\documentclass[10pt,a4paper,onecolumn]{vupreprint}
\usepackage[bindingoffset=0.5cm,textheight=25cm,hdivide={2.5cm,*,3cm}, vdivide={*,24cm,*}]{geometry}
\usepackage{amsmath}
\usepackage{amsfonts}
\usepackage{amssymb}
\usepackage[numbers,square,comma,sort&compress]{natbib}
\usepackage{fancyhdr}
\usepackage{fancybox}
\usepackage{authblk}
\usepackage{siunitx}
\usepackage{booktabs}
\usepackage{multirow}
\usepackage{soul}
\usepackage{doi}
\usepackage{microtype}
\usepackage[pdftex]{graphicx}
\usepackage{upgreek}
\usepackage{textcomp}
\newcommand{\eqnref}[1]{Eqn.~(\ref{#1})}

\newcommand{\figref}[1]{Fig.~\ref{#1}}			
\newcommand{\tabref}[1]{Tab.\ref{#1}}			
\newcommand{\secref}[1]{Section~\ref{#1}}		
\newcommand{\appref}[1]{~\ref{#1}}		

\newcommand{\cannex}{C\textsc{annex}}
\newcommand{\ri}{{\rm i}}						

\newcommand{\x}{\xi}

\renewcommand{\Xi}{\Xi}

\newcommand{\inv}[1]{\frac{1}{#1}}					
\graphicspath{{./}{./Definitions/}{./Definitions/CasimirFigures/}}
 \doipub{not yet available}
 \pacs{quantum vacuum; Casimir pressure; axion; non-Newtonian gravity}
 \journalref{H. Haghmoradi, H. Fischer, A. Bertolini, I. Gali\'c, F. Intravaia, M. Pitschmann, R. Schimpl, and R. I.P. Sedmik}{Physics}{11(3)}{407}{2024}{Force metrology with plane parallel plates:Final design review and outlook}
 \author{{Hamid Haghmoradi}$^{1}${\footnote{Correspondence experiment: hamid.haghmoradi@tuwien.ac.at}} }
 \author{Hauke Fischer$^{1}$\footnote{Conrrespondence DE\&DM theory: hauke.fischer@tuwien.ac.at} }
 \author{Alessandro Bertolini$^{2}$ }
 \author{Ivica Gali\'c$^{1,3}$}
 \author{Francesco Intravaia$^4$\footnote{Correspondence Casimir effect: francesco.intravaia@physik.hu-berlin.de}}
 \author{Mario Pitschmann$^1$}
 \author{Raphael Schimpl$^1$}
 \author{Ren\'e I.P. Sedmik$^1$\footnote{Correspondence experiment and error calculation: rene.sedmik@tuwien.ac.at}}
 \affil{$^1$TU Wien, Atominstitut, Stadionallee 2, 1020 Vienna, Austria}
 \affil{$^2$Nikhef, Science Park, Amsterdam, 1098 XG, The Netherlands}
 \affil{$^3$University of Zagreb, Faculty of Mechanical Engineering and Naval Architecture}
 \affil{$^4$Institut f\"ur Physik, Humboldt Universit\"at zu Berlin, Newtonstra\ss{}e 15, 12489 Berlin, Germany}
\hypersetup{pdfauthor={\theauthors},pdftitle={\thetitle}}
\title{Force metrology with plane parallel plates: Final design review and outlook}
\begin{document}
\maketitle
\begin{abstract}
During the past few decades, abundant evidence for physics beyond the two standard models of particle physics and cosmology was found. Yet, we are tapping into the dark regarding our understanding of the dark sector. For more than a century, open problems related to the nature of the vacuum remain unresolved. Besides the traditional high-energy frontier and cosmology, technological advancement provides complementary access to new physics via high-precision experiments. Among the latter, the Casimir And Non-Newtonian force EXperiment (\cannex{}) has successfully completed its proof-of-principle phase and will soon commence operation. Benefiting from its plane parallel plate geometry, both interfacial and gravity-like forces are maximized, leading to increased sensitivity. A wide range of dark sector forces, Casimir forces in and out of thermal equilibrium, and gravity will be tested. This article describes the final experimental design, its sensitivity, and expected results.
\end{abstract}
\section{Introduction}
\label{sec:intro}
Continuous improvements in measurement methods during the past few decades have unveiled a number of tensions between predictions of the standard models of particle physics (SM) and cosmology ($\Lambda$-CDM) with observations. Since the 1970s, the development and test of the SM have been dominated by collider experiments culminating in the experimental discovery of the Higgs particle. However, further advancement on the high energy frontier appears difficult, as the required technological and financial efforts grow over-proportionally with the gain in energy. Yet there are still 16 orders of magnitude missing between the current \SI{10}{\tera\electronvolt} scale and the Planck scale. Therefore, precision measurements at lower energy have established themselves as an alternative route to test existing theories and to search for physics beyond.

In fact, precision tests have unveiled a growing number of `tensions' in various fields that cannot be explained well on the basis of existing theory.
We can highlight only a few of these here.
For quantum electrodynamics, measurements of the relative gyromagnetic moments $(g-2)/2$ of fermions have revealed values~\cite{Keshavarzi:2021eqa} that differ from theoretical expectations by $2.5\sigma$ for electrons and $4.2\sigma$ for muons, giving a strong signal of either an incomplete understanding of vacuum fluctuation contributions or new physics. Charge radii of the proton and the deuteron have been determined using precision (Lamb-shift) spectroscopic measurements with H and D as well as from electron and muon scattering experiments~(review~\cite{Gao:2021sml}). Even after a recent re-analysis of experimental errors, and new measurements, tensions at the $2\sigma$~\cite{Tiesinga:2021myr} and $3.5\sigma$~\cite{Gao:2021sml} level, respectively, exist between different experiments and between experiments and theory. While QED is still referred to as the `best tested theory', even after $\sim150$ years the question, if whether the electromagnetic energy momentum tensor is traceless or not in materials, remains open~\cite{Burger:2020}. Tensions are also known for other sectors of the SM. For example, the Cabibbo-Kobayashi-Maskawa (CKM) quark mixing matrix of QCD shows increasing signs of non-unitarity (currently $2.2\sigma$~\cite{ParticleDataGroup:2022pth} or up to $2.8\sigma$~\cite{Hardy:2020qwl}), which, if confirmed, would be an indication for beyond SM physics.  
In QCD, the breaking of $CP$ symmetry being suppressed by a factor $10^{-10}$, creates a fine tuning problem that could be resolved~\cite{Peccei:1977hh, Peccei:1977ur} by an additional spontaneously broken `Peccei-Quinn' symmetry leading to the axion as its associated Nambu-Goldstone boson~\cite{Weinberg:1977ma, Wilczek:1977pj}. The latter is constrained strongly but not yet excluded. Another strong motivation for the axion is due to it providing an excellent candidate for dark matter (DM).

DM has a solid basis of evidence, as galaxy rotation curves are measured since the early 20th century~\cite{Zwicky:1933}, and newer probes, such as cosmic microwave background or weak lensing data indicate that a fraction $\Omega_\text{DM}\approx 0.27$~\cite{Planck:2018vyg} of the total mass in our universe can be attributed to DM (see~\cite{ParticleDataGroup:2022pth} for a review). Numerical simulations~\cite{Angulo:2021kes} show that the current large-scale structure of the universe can only be obtained if DM is taken into account, with baryonic matter ($\Omega_b\approx0.05$) playing a sub-leading role.

After the discovery of accelerated expansion 25 years ago~\cite{SupernovaCosmologyProject:1997zqe, SupernovaSearchTeam:1998fmf,SupernovaSearchTeam:1998bnz}, we also have clear indications that by far the largest fraction of the energy/mass content of our universe ($\Omega_\text{DE}\approx0.68$) is due to the existence of what is generically termed 'dark energy' (DE). In general relativity (GR), dark energy can effectively be described in terms of a cosmological constant $\Lambda$ providing the negative pressure necessary to account for an accelerated expansion of our universe. In combination with 'cold' DM this constitutes the cosmological standard model $\Lambda$-CDM. However, as the Hubble constant $H_0$ -- being a measure of expansion -- obtained from data on the cosmic microwave background at large redshift $z$, is at significant tension ($5\sigma$) with the one obtained from local distance ladder measurements at $z<2.36$ and a range of other measurements~\cite{Hu:2023jqc}, speculations arise (among others) if $\Lambda$ is a constant, after all~\cite{Koch:2022cta}. Significant tensions exist not only in measurements of $H_0$ but also for several other parameters of $\Lambda$-CDM~\cite{Perivolaropoulos:2021jda}. Since DE accounts for the largest fraction of the energy/mass content of our universe, the quest for an answer to the question what the dark sector is composed of, receives strong attention. It is currently unknown whether DE and DM are composed of new particles or not, but the answer lies probably beyond the current SM / $\Lambda$-CDM framework.

While amending general relativity by the cosmological constant enables us to describe an accelerated expansion, such a procedure would lead to a severe fine-tuning problem, which is the so-called `(old) cosmological constant problem'~\cite{Sola:2013gha}. This is due to contributions in addition to Einstein's original (bare) cosmological constant, coming from the zero-point energies of all quantum fields (SM fields as well as possible unknown ones) as well as the Higgs potential during its phase transition related to electroweak symmetry breaking~\cite{Martin:2012bt}. Introducing a cutoff at the Planck scale or electroweak unification scale in order to render the zero-point energies finite, these contributions provide values for $\Lambda$ that are 123, respectively, 55 orders of magnitude above the measured value containing all contributions~\cite{Sola:2013gha}. This may suggest that quantum fluctuations of the vacuum do not seem to gravitate~\cite{Padmanabhan:2006cj}, which has cast some doubt on their reality. 
Some have resorted to the rather metaphysical anthropological principle~\cite{Weinberg:1996xe} to explain the `cosmological constant problem'~\cite{Weinberg:1988cp}, while others, just to give an example, have attempted to find explanations in terms of a natural cutoff given by metric feedback at high energies~\cite{Cree:2018mcx}. If there existed additional interactions, cancellations of the zero-point energies of these new fields and the ones of the standard model~\cite{Adler:1995vd} could explain the smallness of $\Lambda$. However, we would be left with a severe fine tuning problem, which adds to the problem of non-gravitation of vacuum fluctuations. By now, a whole host of conceptually distinct approaches has been devised to avoid this problem (see e.g.~\cite{Nobbenhuis:2006yf,Martin:2012bt,Sola:2013gha}) with no final solution. 

While no general consensus has been found on the above tensions, one approach to explain them is to introduce new interactions. The historically very successful approach to search for the associated new particles in colliders, however, has not led to any discoveries so far, for either DM or DE. Indications for weakly interacting massive particles (WIMPs) have not been found at high energies. Lighter particles searched for by recoil experiments have also eluded detection~\footnote{The DAMA collaboration's periodic DM signal is not generally considered to be confirmed at the time of writing.} despite large international efforts. Astronomical observations, on the other hand, may have found indications for sterile neutrinos~\cite{Bulbul:2014sua,Hofmann:2019ihc} and WIMPS~\cite{Barkana:2018lgd}. Indications were also found in long baseline nuclear experiments but are still being discussed. As no clear signs regarding the type or energy range of new interactions have been found, theoreticians have turned to the broad field of effective field theories to give generic predictions that allow experimentalists to narrow down the possibilities for DM and DE models.
Irrespective of the true physical origin, an effective field theory allows to describe and classify the low-energy behavior of the corresponding fundamental theory in a model-independent way. As such, the `Standard Model Extension'~\cite{Colladay:1998fq} covers all possible $CP$($T$)-violating terms that could be added to the SM. Several of these can also be written in terms of bosonic spin-0 or spin 1, scalar, vector, or tensor interactions (and their respective pseudo or axial counterparts) between SM fermions~\cite{Moody:1984ba,Fadeev:2018rfl}. The latter leads to a class of effective potentials that can be tested in a large number of experiments~\cite{Sponar:2020gfr}. For DE, besides modified gravity, variable dark energy models, and black holes, a class of screened scalar fields has been investigated that would describe dynamical `quintessence' scalar fields with an effective potential depending on the local mass density. This local variability permits them to `hide' in denser environments and evade stringent astrophysical bounds while still being able to prevail in low density regions thereby describing DE. However, these models have several free parameters, and only a few, such as the string-inspired dilaton, have a more solid motivation. 

In any case, the cosmological constant problem provides further indications that our understanding of the quantum vacuum may be incomplete. This has been one of many motivations for investigations of the Casimir effect. Being the only known quantum effect causing forces between separated macroscopic objects, experiments have been performed since its prediction in 1948~\cite{Casimir:1948}. Modern experiments starting in the 1990s~\cite{Lamoreaux:1996wh,Mohideen:1998iz,Roy:1999zz} have tested non-trivial boundary dependence~\cite{Tang:2017,Garrett:2018} and lateral forces~\cite{Chen:2002zzb,Chiu:2009fqu}, thin layers~\cite{Lisanti:2005}, dielectric properties~\cite{Chen:2006zz,deMan:2009zz,Torricelli:2010a,Torricelli:2011,Banishev:2012bh,Banishev:2013,Liu:2021ice}, influence on micro-electromechanical elements~\cite{Ardito:2012,Broer:2013bxa}, torque~\cite{Somers:2018}, repulsion~\cite{Lee:2001,Feiler:2008,Munday:2009fgb}, to name just a few topics. Regarding the description of the dielectric properties, especially for the thermal contribution to the Casimir effect, there has been a discussion going on for more than two decades (review:~\cite{Mostepanenko:2021cbf}). 
Specifically, a disagreement between theoretical predictions and experimental results put the focus on the proper account of dissipation in the description of the material optical response. Surprisingly, a simple non-dissipative model provides a better description of several experiments measuring the Casimir interaction between metallic objects. 
At the same time, the same experiments appear to exclude an account of dissipation in terms of the commonly-used Drude model~\cite{Bimonte:2016myg}. A similar issue was noticed for free electrons in semiconductors~\cite{Chen07a}. Within the same context, attention has also been devoted to surface roughness~\cite{Zwol11} and patch potentials~\cite{Behunin:2011gj,Sushkov11} as a possible source for the disagreement between theory and experiment. Other material properties were investigated and in particular, non-locality (spatial dispersion) has attracted attention~\cite{Klimchitskaya:2020qmy} also in relation to thermodynamic inconsistencies, which may appear when the Drude model is adopted for the description of a metal~\cite{Bezerra:2004zz}. However, up to now, all attempts have not reached a unanimous consensus, and more precise experimental data are required to settle the controversy~\cite{Mostepanenko:2021cbf}. More recently, it was pointed out that a non-equilibrium configuration
in which the objects are at different temperatures $T_1$ and $T_2$ can serve as an additional benchmark of the theoretical framework surrounding the Casimir effect~\cite{Klimchitskaya:2019nzu}.
In this case, an additional contribution to the interaction, anti-symmetric under the exchange $T_1\leftrightarrow T_2$, has been predicted. Still, this contribution has not yet been quantitatively confirmed in a Casimir experiment.

Experimentally, precision Casimir experiments have also been used to set limits on new interactions~\cite{Bordag:1999gx,Mostepanenko:2001fx,Decca:2003td,Decca:2005qz,Decca:2007jq,Mostepanenko:2008,Masuda:2009vu,Bezerra:2010pq,Sushkov:2011md,Klimchitskaya:2013rwd,Bezerra:2014ona,Chen:2014oda,Klimchitskaya:2017cnn} at small separation $a$, as proposed four decades ago~\cite{Kuzmin:1982}. However, the sensitivity is limited~\cite{Behunin:2013qba}, as one of the strongest uncertainties in such measurements comes from local surface charges that are hard to quantify and control~\cite{Behunin:2011gj,Behunin:2013qba}. These uncertainties can be mostly avoided by using the `iso-electronic' technique~\cite{Decca:2005qz,Bimonte:2016myg,Wang:2016ngs} (allowing only relative measurements) or by placing an electrostatic shield between the test objects, leading to the Cavendish configuration that has extensively been used in torsion balance experiments~\cite{Adelberger:1990xq,Hoyle:2004cw,Adelberger:2006dh,Hammond:2007jm,Schlamminger:2007ht,Heckel:2008hw,Geraci:2008hb,Hoedl:2011zz,Heckel:2013ina,Terrano:2015sna,Tan:2016vwu,Tan:2020vpf,Lee:2020zjt,Zhao:2021anp} to measure gravity-like interactions. However, a shield between the interacting objects precludes the measurement of Casimir forces and DE screened scalar fields. Another common disadvantage of most existing precision force experiments in either configuration is that they use curved surfaces of some radius $R_s$. Depending on the distance dependence of the investigated interaction, the effective surface area generating the force is thereby dramatically reduced from $A=R_s^2\pi$ to $A_\text{eff}\approx\pi R_s a$~\cite{Sedmik:2013vna,vanZwol:2009} with $a\ll R_s$. Therefore, one looses a factor $\eta_\text{eff}\equiv a/R_s=10^{-2}\ldots 10^{-4}$ in force sensitivity~\cite{Sedmik:2021iaw}. This problem is maximally avoided for plane parallel plates, where $\eta_\text{eff}=1$. The downside is that one has to measure and control parallelism and use perfectly flat surfaces, which introduces significant technical difficulties. Previous attempts to measure Casimir force gradients between parallel plates~\cite{Bressi:2002fr,Antonini:2008bb} have suffered from electrostatic and other unresolved offsets, for which the results included a free fit parameter.

The Casimir And Non-Newtonian force EXperiment (\cannex{}) has been designed from the onset to perform measurements between macroscopic plane parallel plates~\cite{Almasi:2015zpa}. After a first proof of principle~\cite{Sedmik:2018kqt}, we continuously updated the design~\cite{Sedmik:2020cfj,Sedmik:2021iaw} to characterize, attenuate, or actively control all relevant disturbances. The setup allows synchronous measurements of the pressure and pressure gradient with nominal sensitivities of \SI{1}{\nano\newton/\metre^2} and \SI{1}{\milli\newton/\metre^3}, respectively, in both Cavendish and Casimir configuration, in the distance regime \SI{3}{\micro\metre}--\SI{30}{\micro\metre}. Recently, we selected the Conrad Observatory (COBS), a geoseismic and geomagnetic surveillance station inside a tunnel system in the Alps as a location. The seismic and thermal stability there will reduce errors and technical requirements of isolation systems for \cannex{}. Operations at COBS are expected to commence in the summer of 2024.

In this article, we present the final design and its predicted performance in Cavendish and interfacial (Casimir configuration) in \secref{sec:setup}. Subsequently, we update our recent error budget~\cite{Sedmik:2021iaw} taking into account actual device specifications and preliminary noise measurements. We then give an update on prospects for measurements of in- and out-of-equilibrium Casimir forces~\cite{Klimchitskaya:2019nzu} in \secref{sec:casimir}. Finally, we present updated prospects for DE screened scalar field limits~\cite{Almasi:2015zpa,Sedmik:2021iaw,Fischer:2023koa} in \secref{sec:de}, based on fully-consistent numerical calculations taking into account the experimental and theoretical uncertainties, and close with a short outlook in \secref{sec:discussion}.

\section{Experimental Design}
\label{sec:setup}
\begin{figure}[hb!]
 \centering
  \includegraphics[width=1\textwidth]{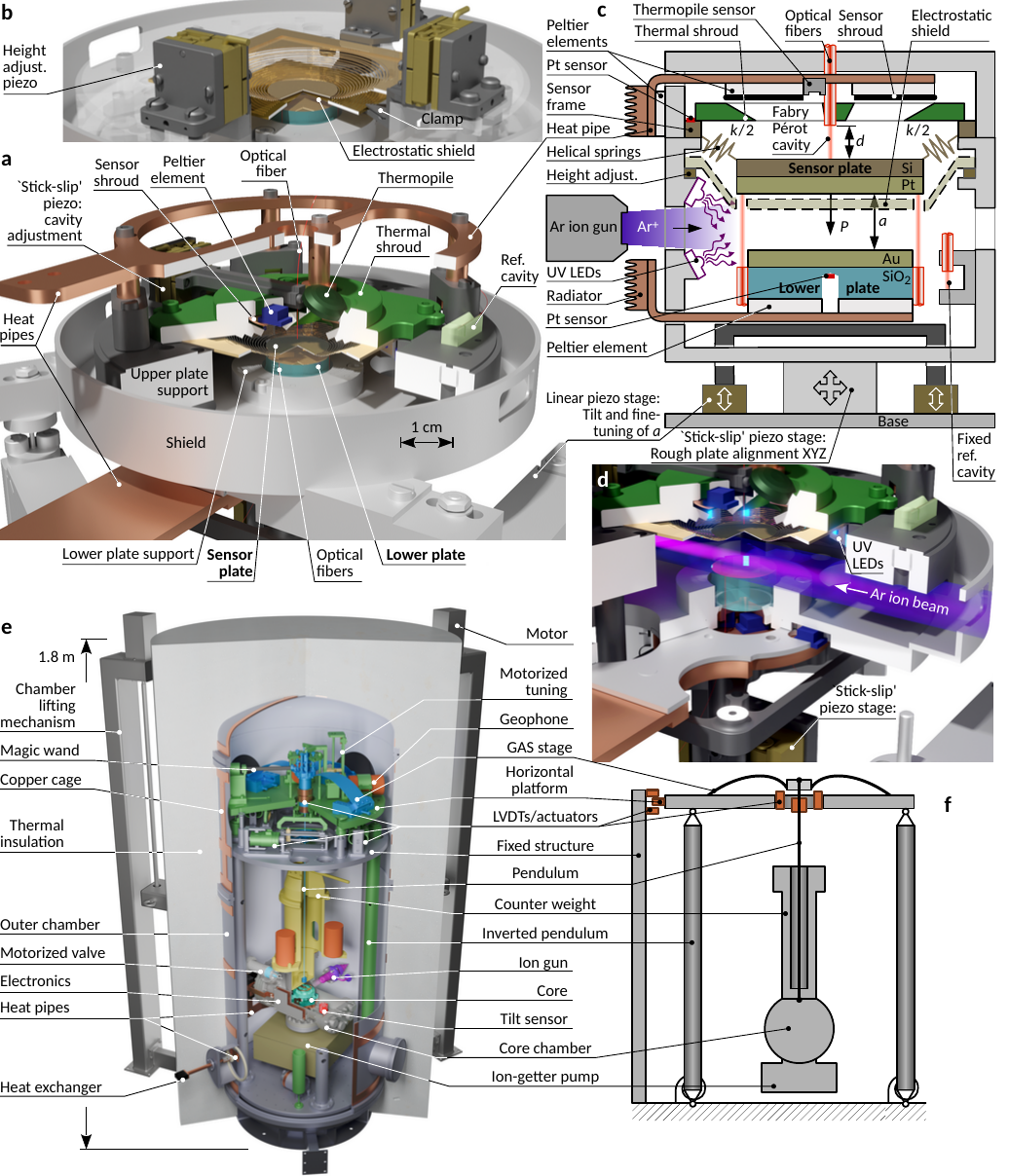}
 \caption{The \cannex{} setup. \textbf{a)} Simplified cut view of the actual core design in interfacial configuration. \textbf{b)} Simplified focus view of the Cavendish configuration with the electrostatic shield and associated adjustment stages. \textbf{c)} Schematic representation of the core including all elements and configurations. \textbf{d)} View of the core with the translator stages in their upper position, in which the ion tunnels are opened for Ar ion cleaning and UV irradiation. \textbf{e)} Cut view of the complete setup. \textbf{f)} Schematic view of the seismic attenuation system (SAS).\label{fig:setup_overview}}
 \end{figure}

\cannex{} is a metrological setup designed to synchronously measure forces and force gradients between plane parallel plates at separations between \SI{3}{\micro\metre} and \SI{30}{\micro\metre} in interfacial and Cavendish configuration. Force (gradients) are measured by interferometrically detecting (see \secref{sec:setup:omsystem}) the movement of a mass-spring system consisting of a `sensor' plate and a set of helical springs. Forces onto the sensor plate are sourced by a second fixed `lower' parallel plate, as shown in \figref{fig:setup_overview}d. Since this mechanical detection system is highly sensitive to mechanical vibrations, surface charges, and thermal changes we have included countermeasures for all of these disturbances in terms of active control and attenuation systems into the design described in detail in sections \ref{sec:setup:sas}, \ref{sec:setup:chargecomp}, and \ref{sec:setup:kpfm}, respectively. In the following, we give an overview of the setup.

The setup's `core' contains the actual measuring device. Here, the lower plate, made of silica glass, is mounted in a fitting (light gray in \figref{fig:setup_overview}a--d) that isolates it thermally and electrically from the rest of the setup. The fitting is supported by three linear piezo transducers with a range of \SI{100}{\micro\metre} allowing us to fine-tune the parallelism and separation between the plates. Thermal control of the lower plate can be achieved via Peltier elements (PE) below it and a platinum sensor at its center. Attached to the side of the lower late are three optical fibers used to measure plate separation and tilt (see \secref{sec:setup:omsystem}).

The force sensor is fabricated from a silicon single crystal {(Norcada Inc.)} and placed directly above the lower plate. Its position can be adapted by a three-axis drift-free stick-slip stage {(Smaract SLC-1720)} supporting the entire upper part of the core. The sensor's frame is connected to a massive support structure (middle gray in \figref{fig:setup_overview}a, c, d) that is thermally controlled by distributed PEs, and electrically grounded. The support carries a thermal shroud (green) allowing for non-contact thermal control of the sensor and its springs. 

Sensor movements are detected via an optical fiber placed above its center (see \secref{sec:setup:omsystem}). The fiber is attached to a drift-free stick-slip piezo transducer allowing us to adjust the cavity size.
Similarly, the separation $a$ between the lower and upper plate is monitored by three interferometers arranged around the rim of the lower plate. The fibers' end faces are polished optically together with the lower plate in order for them to be at exactly the same height. 
\cannex{} implements three different configurations. In the first -- \emph{interfacial} -- configuration, the sensor plate directly faces the top surface of the lower plate. In the second -- \emph{Cavendish} -- configuration, we add a gold-coated silicon nitride membrane acting as an electrostatic shield (ESS) between the two plates. The ESS is held by three stick-slip piezos (see \figref{fig:setup_overview}b) to change its height and orientation. Despite its large area (\SI{1}{\centi\metre^2}) and small thickness ($<\SI{1}{\micro\metre}$), the ESS has an extremely low hang-through under gravity of about \SI{1}{\micro \metre}. Three pinholes in the ESS allow the lower plate's interferometers to operate both through the ESS (to measure $a$) and when shifted slightly to the side, to monitor the separation between ESS and lower plate. This mechanism allows us to unambiguously determine and control the relative position of all three plates with respect to each other. In the third configuration discussed in \secref{sec:setup:kpfm}, which is only used for surface characterization, either the sensor and the shroud, or the lower plate and its fitting are replaced by a Kelvin probe setup able to scan the surface potential and topology of the entire surface area of the remaining plate.

The core assembly, \figref{fig:setup_overview}a, is enclosed inside an ultra-high vacuum (UHV) `core' chamber. This chamber can be evacuated down to a pressure of \SI{e-9}{\milli \bar} by using an ion-getter pump or be filled with up to \SI{500}{\milli\bar} of Xe gas for measurements of screened DE interactions~\cite{Brax:2010xx}. On the outside of the core chamber wall, the core electronics are placed on a copper plate that allows generated heat to be guided away without mechanical contact to the outside of the outer chamber (shown partially in \figref{fig:setup_overview}e). A similar but independent mechanism exists for the heat pipes emerging from the core itself. Details on these systems are given in \secref{sec:setup:temp}.
The core chamber is suspended on a 6-axis seismic attenuation system (SAS) shown in \figref{fig:setup_overview} e and f. The SAS comprises an inverted pendulum (green) for horizontal isolation, a geometric anti-spring (GAS) filter (blue) for vertical isolation, and a mass tower (yellow) improving tilt isolation, as described in more detail below. Additionally, a hollow silicon carbide rod known as a compensation wand (magic wand) is connected to the tip of the GAS filter to improve the attenuation performance \cite{Stochino:2007zz}. Vertical and horizontal positions of the SAS can be sensed by linear variable differential transformer (LVDT) sensors, and controlled by motorized pre-tension springs. In addition, the dynamical behavior in translational degrees of freedom can be influenced by voice coil actuators. For higher sensitivity at intermediate frequencies, geophones are used to monitor all but the vertical rotation degree of freedom. Inverted pendulums (IPs) support the base plate of the GAS filter, thereby combining vertical and horizontal attenuation systems. 

The entire SAS with the core is enclosed in an `outer' vacuum chamber at $10^{-6}\,$\si{\milli \bar} providing further isolation against sound, thermal, and other environmental disturbances. This chamber is mechanically decoupled from the SAS, to ensure deformations due to pressure differences will not influence the performance of the SAS. The outer chamber is fitted with a dense grid of copper bars and a \SI{25}{\centi\metre} insulation layer to reduce temperature gradients on the chamber wall. We use PEs on the mentioned copper bars to control the chamber temperature with a precision of $\sim4\,$\si{\milli\kelvin} in order to isolate \cannex{}'s temperature. Eventually, the chamber includes several exterior mechanisms (not shown) to open it, and to extract the core with minimum mechanical input to the sensor. The entire setup is placed inside an ISO class 7 cleanroom inside the tunnels at COBS. 

\subsection{Seismic Attenuation}
\label{sec:setup:sas}

Seismic vibrations present a formidable impediment to the fidelity of small-distance metrology setups, necessitating a comprehensive understanding of their impact on the respective measurements. \cannex{} uses non-linear mechanical elements developed for gravitational wave detectors~\cite{Stochino:2009zz,Blom:2015fna,Heijningen:2019jmd}. For vertical isolation, a geometric anti-spring (GAS) filter~\cite{Cella:2004bn} (blue in \figref{fig:setup_overview}e) provides 40\,dB/decade attenuation from $\sim 100\,$mHz. We employ so-called `magic wands '~\cite{Stochino:2007zz} to augment filter performance at low frequencies and near the sensor resonances. Horizontal isolation is achieved by inverted pendula~\cite{Takamori:2007zz} (green) carrying the GAS filter and a regular pendulum suspending the core chamber. The tilt of the core chamber around the horizontal axes is attenuated by the core chamber being mounted on the pendulum close to its center of gravity. The latter is raised to the hinge point by means of a massive tower (yellow), which reduces the tilt resonance frequency.

\begin{figure}[!hb]
    \centering
    \includegraphics[width=\textwidth]{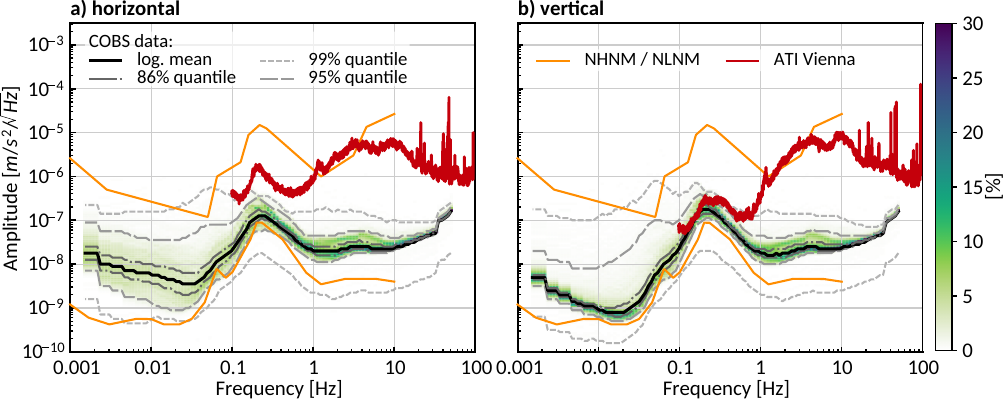}
    \caption{Seismic background in \textbf{a)} vertical and \textbf{b)} horizontal (eastern) direction. Data were recorded at COBS using calibrated STS-2 seismometers between 2023-04-30 and 2023-06-30, including 4 earthquakes of magnitude up to 3.3. The black line is the logarithmic mean of the data, while the dashed gray lines represent the quantiles obtained from histograms at each frequency. The green and blue color encodes the probability. In comparison, the seismic background at the Atominstitut (ATI) in Vienna, recorded using calibrated Sercel L4C geophones during 24 hours starting 2018-09-28 is significantly higher due to a nearby highway, subway, and in-house noise sources. For reference, we give Peterson's new high and low noise models~\cite{Peterson:1993a} (NHNM and NLNM, respectively).}
    \label{fig:seismic_background}
\end{figure}

Previously, the Atominstitut of TU Wien was considered as the location for \cannex{}~\cite{Sedmik:2021iaw}. We have identified a more suitable location in the underground laboratory of the Conrad observatory, roughly \SI{50}{\kilo \metre} southwest of Vienna. Seismic spectra have been recorded at both locations and are shown in \figref{fig:seismic_background}. As we will discuss in \secref{sec:error:seismic} below, a one-staged passive SAS at COBS already fulfills all requirements for the targeted error level while at ATI, a two-stage isolation system would be required~\cite{Sedmik:2021iaw}. In the following, we describe the final system, which is similar to already realized systems in the literature~\cite{Blom:2015fna,Heijningen:2019jmd}.

\begin{figure}[!ht]
 \centering
 \includegraphics[width=\textwidth]{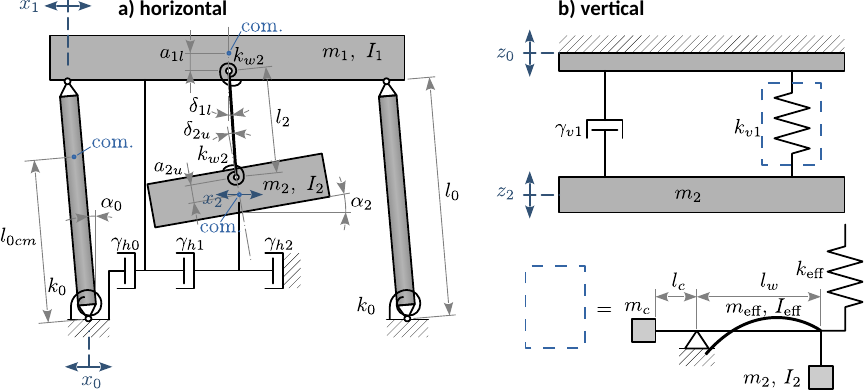}
 \caption{Model for the dynamical behaviour of the \cannex{} SAS in \textbf{a)} horizontal and \textbf{b)} vertical direction. The centers of mass are indicated by `com.'.\label{fig:setup:sas:model}}
\end{figure}
The principle of GAS filters~\cite{Cella:2004bn} and inverted pendula~\cite{Takamori:2007zz} relies on the instability of non-linear mechanical systems at which the stable operating point splits into two distinct solutions. Such points lead (theoretically) to zero resonance frequency and thereby ideal isolation. In practice, internal damping and creep set limits on the achievable minimum resonance frequencies. We can describe our SAS by the model shown in \figref{fig:setup:sas:model}. The respective small-signal Lagrange functions $\mathcal{L}_D$ for horizontal ($D=h$) and vertical ($D=v$) directions are given by
\begin{align}
\mathcal{L}_h=&\mathcal{T}_h&&\hspace{-2.5 ex}-\mathcal{V}_h\\
\text{with }& \mathcal{T}_h&&\hspace{-2.5 ex}=\inv{2} \left[I_0 (\partial_t\alpha_0)^2 + I_2 (\partial_t\alpha_2)^2 + m_1 (\partial_t x_1)^2 + m_2 (\partial_t x_2)^2 + m_0  (\partial_t x_c)^2\right]\,,&\hfill\nonumber\\
 & \mathcal{V}_h&&\hspace{-2.5 ex}=\inv{2} \left[k_0 (x_1 - x_0)^2 + k_{w2} (\delta_{1}^2 + \delta_{2u}^2)\right] + m_0 g y_0 + m_1 g y_1 + m_2 g y_2\,,\hfill\nonumber\\
 & \mathcal{R}_h&&\hspace{-2.5 ex}=\inv{2}\left[\gamma_{2h} (\partial_tx_2 - \partial_tx_0)^2 + \gamma_{1h} (\partial_tx_2 - \partial_t x_1)^2 + \gamma_{0h} (\partial_tx_1 - \partial_t x_0)^2\right]\,,\nonumber\\
& x_c&&\hspace{-2.5 ex}= x_0\left(1-\frac{l_{0cm}}{l_0}\right) +x_1\frac{l_{0cm}}{l_0} ,\nonumber\\
& y_0 &&\hspace{-2.5 ex}= -\inv{2} \alpha_0^2 l_{0cm}\,,\quad y_1 = -\inv{2} \alpha_0^2 l_0\,,\quad y_2 = y1 + \inv{2} (\delta_{1}^2 l_2 + \alpha_2^2 a_{2u})\,,\quad\nonumber\\ 
& \delta_1 &&\hspace{-2.5 ex}= \frac{x_2-x_1-a_{2u} \alpha_2}{l_2}\,,\quad \delta_{2u}=\alpha_2-\delta_1\,,\quad \alpha_0 =\frac{x_1-x_0}{l_0}\,,\quad \partial_t\alpha_0 =\frac{\partial_tx_1 - \partial_tx_0}{l_0}\,.\hfill\nonumber\\
\mathcal{L}_v=&\mathcal{T}_v &&\hspace{-2.5 ex}-\mathcal{V}_v\,,\\
 \text{with }&\mathcal{T}_v &&\hspace{-2.5 ex}=\inv{2}\left[ m_2 (\partial_t z_2)^2 + m_c\left(\frac{l_c}{l_w} (\partial_t z_2 - \partial_t z_0)\right)^2\right]\,,\nonumber\\
 &\mathcal{V}_v &&\hspace{-2.5 ex}=\inv{2} k_{v1} (z_2 - z_0)^2 \,,\quad  \mathcal{R}_v=\inv{2}\gamma_1 (\partial_tz_2 - \partial_t z_0)^2\,.\nonumber
\end{align}
Here, in $\mathcal{T}_h$ the first two terms describe angular kinetic energy of the top and payload masses $m_1$ and $m_2$, respectively. The last three terms are the linear kinetic energies of the top, payload, and inverted pendulum ($m_0$) masses. Potential energies for the inverted pendulum tilt and wire tilt are given by the first two terms in $\mathcal{V}_h$, while the last two terms regard the change in the absolute height of $m_1$ and $m_2$ due to rotary (sidewards) movements of the inverted pendulum and pendulum, respectively. Viscous damping between all parts is considered by the Rayleigh dissipation terms in $\mathcal{R}_h$, while internal friction is added \emph{ad hoc} by adding to the effective values $k_i=\omega_i^2 m_i$ representing the spring constant of system $i$ a factor $(1+\ri\phi)$ with $\phi<1$~\cite{Stochino:2007zz}, not shown here for brevity.
Similarly, for the vertical direction, we have the linear kinetic energy of the payload mass and wand counter weight $\mu_1$ in $\mathcal{T}_v$. Deformation of the GAS filter gives a contribution to the potential energy $\mathcal{V}_v$, while viscous damping between the top stage and the payload contributes to the damping term $\mathcal{R}_v$. $y_1$ approximates the vertical shift of $m_1$ for small $\alpha_0$. Consequently, $y_2$ is the vertical shift of $m_2$ due to the combined action of the pendulum and inverted pendulum. The $I_x$ denote the moments of inertia of the inverted pendula ($x=0$), the upper platform ($x=1$), the payload (core chamber, $x=2$), and the GAS springs ($x=\text{eff}$) obtained numerically from CAD software. $\gamma_x$ denote damping coefficients, $k_x$ are (effective) elastic constants, and $m_x$ are masses as defined in \figref{fig:setup:sas:model}

\begin{figure}[!ht]
 \centering
 \includegraphics[width=\textwidth]{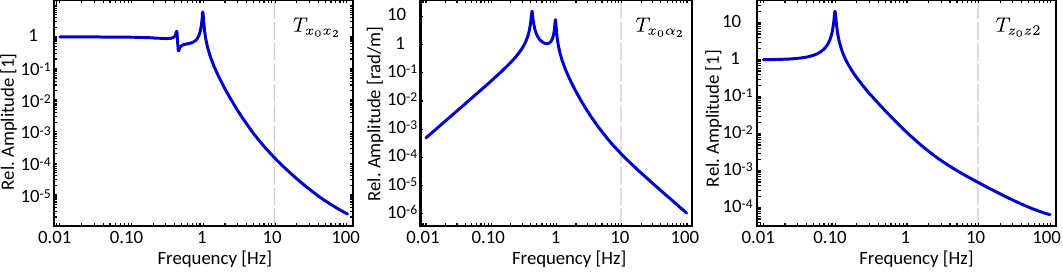}
 \caption{Transfer functions of the passive \cannex{} SAS from horizontal and vertical vibrations to core movement and tilt, respectively, as indicated in the figure. The vertical dashed lines indicate the position of the sensor resonance.\label{fig:setup:sas:performance}}
\end{figure}

The Euler-Lagrange equations giving the dynamical behavior of the system are then
\begin{align}
 \frac{\partial \mathcal{L}}{\partial u}-\frac{\partial}{\partial t}\frac{\partial\mathcal{L}}{\partial (\partial_t u)}-\frac{\partial \mathcal{R}}{\partial (\partial_tu)}=0\,,\label{eq:euler-lagrange}
\end{align}
for $u=x_i,\,z_i\,,\alpha_i$. \eqnref{eq:euler-lagrange} can be resolved for the transfer functions $T_{x0x2}\equiv\frac{X_2}{X_0}$, $T_{x0\alpha_2}\equiv\frac{\alpha_2}{X_0}$, and $T_{z0z2}\equiv\frac{Z_2}{Z_0}$ for horizontal, tilt, and vertical degrees of freedom. We have optimized the system's parameters  with respect to low resonance amplitude and maximum attenuation around the sensor resonance frequency $f_0=\SI{9.8}{\hertz}$, resulting in  the responses shown in \figref{fig:setup:sas:performance}. We obtain a vibration suppression of about 77\,dB and 66\,dB in horizontal (both axes) and vertical direction, respectively at $f_0$, which suffices to achieve the targeted sensitivity under all circumstances, as discussed further in \secref{sec:error}.

The current design includes linear variable differential transformers (LVDT) combined with voice coils for force feedback on horizontal and vertical degrees of freedom as well as geophones. These sensors and actuators shall be used to reduce the amplitude of resonances and improve the overall performance as described for similar systems~\cite{Beker:2014zba}. The design of the feedback system is still in progress.

\subsection{Surface Charge Cancellation}
\label{sec:setup:chargecomp}
Electrostatic patch potentials due to local variations in the Fermi potential of surfaces, chemical impurities, and charge accumulation~\cite{Rossi:1992} are a major nuisance in all interfacial force experiments~\cite{Behunin:2013qba,Garrett:2020acf,Robertson:2006}. Over the years, several methods to characterize and compute these forces (gradients) have been developed~\cite{Speake:2003zz,Kim:2009mr,Behunin:2011gj,Fosco:2013xsa,Garrett:2020acf}. Recently, a new experimental approach was presented to reduce surface charges \emph{in situ}~\cite{Liu:2020zph}. UV irradiation can be used to dissociate larger molecules and extract electrons from surfaces. The residual impurities can then easily be removed by a variant of plasma cleaning using a low-energetic beam of Ar ions. After the process, the surfaces have been demonstrated to exhibit strongly reduced local variations in the potential and a low overall force minimizing potential. 

In \cannex{} we implement a dual strategy. Firstly, in all configurations (interfacial and Cavendish), active surface cleaning using an Ar ion source and UV irradiation, are possible without the need to break the vacuum. This is enabled by the vertical translator stage of the upper plate being able to lift the sensor to a distance of $\sim\SI{5}{\milli\metre}$ above the lower plate and simultaneously open a window in the shield to clear the path for an Ar ion beam (see \figref{fig:setup_overview}d). Using 8 high-power LEDs viewing the gap between the plates, we can apply UV irradiation at $275\,$nm wavelength with up to 2\,W in short pulses (visible in \figref{fig:setup_overview}c and d).\\*
Secondly, the performance of the cleaning procedure can be monitored \emph{in situ} using a custom-built Kelvin probe setup (see \secref{sec:setup:kpfm}) mounted in place of the force sensor. Once the distribution, stability, and amplitude of the surface potentials after cleaning, and intermittent exposure to air~\cite{Rossi:1992} have been determined on both interacting surfaces, the regular force measuring configuration will be restored to perform exactly the same cleaning procedures as before.

Besides patch potentials, two opposing surfaces will differ in their absolute potential even if grounded together, due to contact potentials. To cancel these, we use an active homodyne compensation method that was successfully applied in recent measurements of Casimir forces~\cite{deMan:2009,Sedmik:2013,Sedmik:2023ine} and in the proof of principle for \cannex{}~\cite{Sedmik:2018kqt}. The method is similar to amplitude modulation Kelvin-probe force microscopy (see below), and relies on a small electrostatic excitation $v_\text{AC}(t)=V_\text{ac}\sin \omega_\text{AC} t$ being applied to the lower plate resulting in signals at frequencies $\omega_\text{AC}$ and $2\omega_\text{AC}$ whose amplitudes are measured using a lock-in amplifier. The prior signal is then used to drive a feedback circuit that applies an additional potential $V_\text{DC}$ to the plate, thereby driving $V_\text{DC}-V_0$ to zero~\cite{deMan:2009,Sedmik:2018kqt} with high accuracy. The signal at $2\omega_\text{AC}$ can be used to independently measure the surface separation electrostatically or, what is not required in \cannex{} due to the optical method, to perform an independent measurement of the mechanical properties of the sensor. Importantly, all potentials are applied to the lower plate, while the sensor and all other parts are kept on ground potential. Note also that all surfaces and contacts, except for isolating spacers, are coated by gold to exhibit the same absolute surface potential. The real potentials applied to both plates are measured at all times using a calibrated \emph{in-situ} electrometer amplifier.
%
\subsection{Temperature control}
\label{sec:setup:temp}
\begin{figure}[!ht]
    \centering
    \includegraphics[width=1\linewidth]{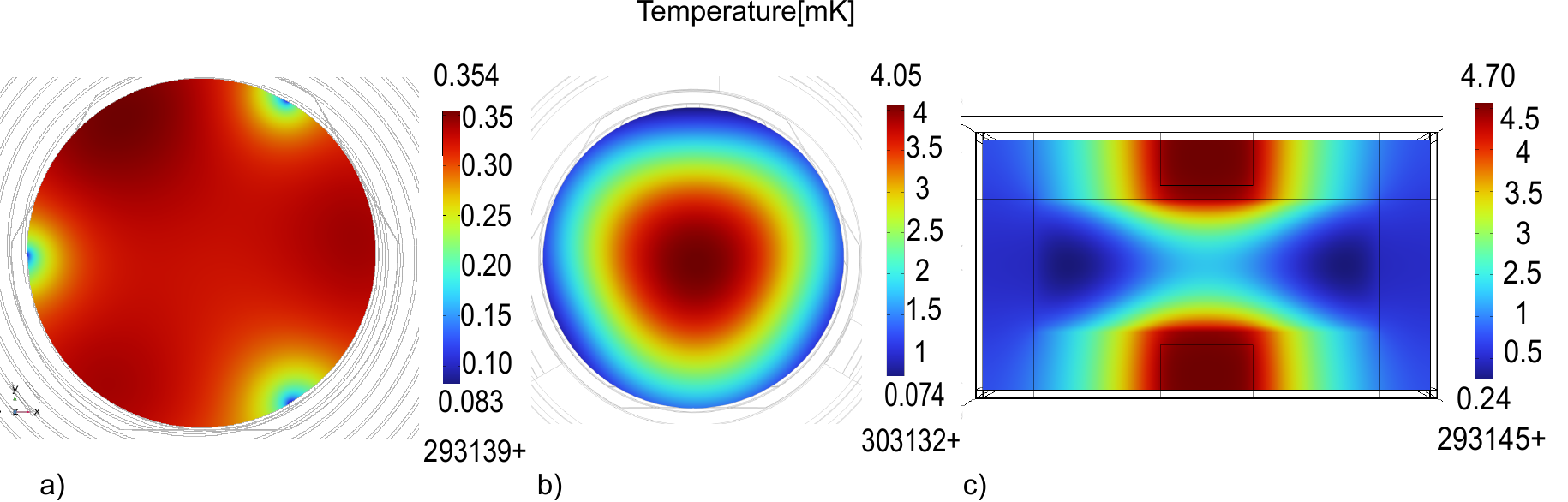}
    \caption{Results of a FEM analysis of thermal distributions. \textbf{a)} sensor plate \textbf{b)} lower plate in the core for a nominal difference of \SI{10}{\celsius} between the two plates. \textbf{c)} chamber wall grid cell (see text). Here, only the relative deviations on the parts are accurate, while the absolute temperatures may contain small offsets due to fixed power input instead of feedback control in the numerical computations.}
    \label{fig:Thermal_Design}
\end{figure}

%
According to the error budget described in \cite{Sedmik:2021iaw} achieving the targeted error levels of $\SI{1}{\nano \newton}/\si{\metre}^2$ and $\SI{1}{\milli \newton}/\si{\metre}^3$ is only possible if thermal stability of the sensor and optical cavities of \SI{0.1}{\milli\kelvin} is guaranteed. In order to comply with this requirement, both plates have an independent thermal control system, responsible for providing the desired thermal stability. We use calibrated custom-made low-noise controllers with 24 bit converters. The lower plate's thermal system is based on thermal conduction and consists of several PEs located below a copper plate attached to the bottom of the lower plate, and a platinum sensor situated in a hole at the center of the lower plate body, to read the temperature as close as possible to the top surface of the plate. The copper plate is insulated thermally from other parts by reflective coatings and mechanically by a gap to suppress heat transfer via radiation and conduction, respectively, between the lower plate and other parts of the core. The lower plate itself is clamped down onto the copper plate and into its fitting using spring-loaded ruby balls, which minimizes mechanical contact area and limits heat loss towards the sides. The sensor plate's temperature is stabilized with a combination of a contact and a non-contact feedback loop. The contact loop consists of several platinum sensors and PE combinations on a copper plate between the upper plate support (\figref{fig:setup_overview}a) and the thermal shroud on top of it, thereby controlling the sensor frame. The central disk of the sensor is connected to its frame only via the long and thin spring arms, for which thermal conduction plays a minor role. On the other hand, the sensor plate exchanges radiation with the lower plate. Despite highly reflective metal coatings on both sides, heat will be transferred between the two plates as soon as they have different temperature setpoints. As the function of the sensor precludes any mechanical contact with its center, the only option to stabilize the upper plate's temperature is via radiation. For this purpose, a blackened copper ring being controlled in temperature by a separate circuit, is placed inside the shroud such that it is visible from the sensor plate's surface. This ring counteracts radiation heating or cooling of the sensor plate by the lower plate. Importantly, we optimized the view factor to the sensor's springs such that they are minimally influenced by radiation from either the ring or the lower plate. Another opening in the shroud allows a thermopile to view the sensor plate and monitor its temperature with $<\SI{0.1}{\milli\celsius}$ precision. This input is used to control the temperature of the ring and in consequence the temperature of the sensor disk. 
A finite element method (FEM) study has been conducted using COMSOL Multiphysics to examine the temperature and respective gradients in all parts of the core. \figref{fig:Thermal_Design} a and b show preliminary results of this study for the temperature distributions of both plates for a temperature setpoint of the lower plate being \SI{10}{\celsius} higher than the ambient temperature. The deviation on the upper plate with respect to the setpoint (\SI{293.130}{\kelvin}) is kept below \SI{0.27}{\milli\kelvin} while on the lower plate, the deviation reaches \SI{3.31}{\milli\kelvin}. 
Proper operation of a PE requires the side opposite to the controlled surface to be connected to a heatsink. 
Therefore, all PEs atop the upper plate, support and shroud are connected via vertical copper columns to the large circular heatpipe on top connected via flexible parts (not shown) to a thermal feedthrough at the back chamber wall. Below the lower plate, the non-control side of the PEs is connected to a heat pipe (shown on the lower left of \figref{fig:setup_overview}d), which leads to a radiator permitting contact-less heat transmission between the core and a feedthrough at the back chamber wall of the core. Similarly, there are two radiators between the core chamber and outer chamber that contactlessly exchange heat with their respective counterpart, partially visible in \figref{fig:setup_overview}e. These radiators on the inner side of the outer walls are connected to heat pipes leading through the outer wall to a thermal controller regulating the heat pipe temperature and effectively releasing excess heat via a heat exchanger to the environment. The radiators themselves are interleaved comb-like structures with a large area overlapping between the interacting parts, blackened on the inside and reflective on the outside. Optimization and testing of these structures is still in progress.


The strong dependency of the working point of non-linear mechanical elements of the SAS on thermal variations~\cite{Cella:2004bn} makes it crucial to actively stabilize the temperature of the outer chamber wall as well to within $\sim5\,$mK. This requirement is not changed by our DC-feedback with the pre-tension springs. In this situation, the low amount of power produced by \cannex{}'s SAS, which still could be a critical thermal disturbance, will be dissipated via radiation interaction with the wall.  At the setup's location at COBS, the ambient temperature changes by much less than \SI{0.1}{\celsius} per day with an average of roughly \SI{10}{\celsius}, for which we expect little exterior thermal fluctuations. To keep the chamber close to the setpoint ($\sim\SI{290}{\kelvin}$), we add \SI{25}{\centi \metre} of passive thermal isolation around the entire chamber. Below the isolation, the entire chamber wall is covered with a dense grid of \SI{5}{\milli\metre} thick copper bars to improve heat conduction on the walls. On top of the copper bars, we add 50 independent calibrated control units consisting each of two PEs and two platinum resistors. In \figref{fig:Thermal_Design}c we show the results of a FEM study of the resulting temperature distribution on the inside of the chamber wall for one representative unit cell of the gridded chamber wall.

\subsection{AFM/KPFM Setup}
\label{sec:setup:kpfm}
\begin{figure}[!ht]
    \centering
    \includegraphics{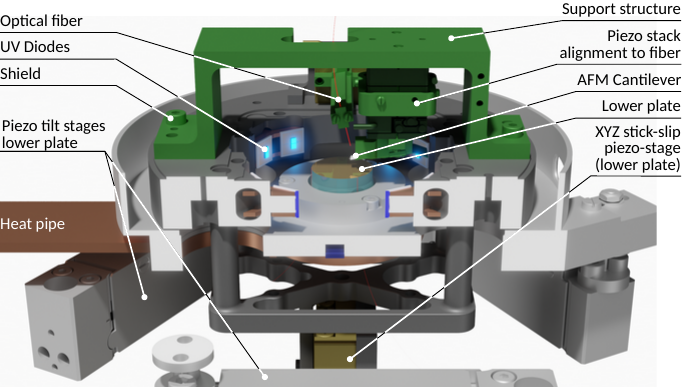}
    \caption{Rendering of the AFM/KPFM configuration for measurements of potentials and topology of the lower plate.\label{fig:kpfm}}
\end{figure}
In section \ref{sec:setup:chargecomp} we discussed the setup for surface charge cancellation by the combined action of an Ar-ion beam and UV irradiation cleaning. To ensure the consistency and performance of these methods, to investigate the long-time evolution of the surface potentials, and to measure the influence of exposure of the setup to the atmosphere \cite{Turetta:2021} (which is inevitable while working on it) we add a Kelvin-probe force microscope (KPFM) to the setup.
The KPFM has been designed to offer two configurations. In the first one shown in \figref{fig:kpfm}, the surface charge distribution on the lower plate can be investigated \emph{in situ} before and after cleaning. 
The sensor plate and shroud (see \figref{fig:setup_overview}) of the force sensing setup, are replaced by a u-shaped structure (green) carrying an AFM cantilever. The optical fiber, which is normally used for measurements on the sensor plate, is remounted at an angle to detect the movement of the cantilever interferometrically, similar as demonstrated previously~\cite{Chavan:2010,Sedmik:2023ine}. 
In order to align the fiber with the cantilever tip \emph{in situ}, we use a stack of horizontal stick-slip translators. The scanning motion of the tip and vertical coarse alignment is carried out by the movement of the 3-axis stick-slip piezo translator stack (golden, at the bottom in \figref{fig:kpfm}). Note that these stages have a range of more than 12\,mm, for which they can be used to investigate the entire area of the lower plate with the KPFM. Because of the comparably large surface separation in \cannex{}, only patches of size $\lambda_p\gtrsim a/10$ are of interest~\cite{Behunin:2014tsa}, with $a$ being the separation between the plates.
Therefore the tip of the cantilever is chosen to be of spherical shape with a diameter of a few \si{\micro\metre}. Using a common sharp tip the same setup can also be used to characterize roughness (and potentials) on all scales with lateral resolution $<10\,$nm. In the vertical direction, we implement a common tapping mode method, where the height adaptation with $0.2\,$nm resolution is enabled by the three linear piezo-electric stages normally used for tilt adjustment of the lower plate.

In the second configuration (not shown here), the KPFM is turned upside down and the lower plate is replaced by the cantilever holder, allowing an \emph{in-situ} measurement of the upper plate's lower surface or the electrostatic shield's potential distribution in the same way as described above. With these two configurations, we can achieve a complete characterization of all surfaces that can then be used to compute the patch potential contribution to the measurements based on actual data instead of statistics.

It has been demonstrated that frequency-modulated (FM-) KPFM is able to achieve higher resolutions than amplitude-modulated (AM-) KPFM, as artifacts caused by the capacitance of the cantilever are more prominent in AM-mode operation~\cite{Zerweck:2005, Axt:2018, Ma:2014}. On the other hand, lower bias voltages in AM KPFM reduce the distance dependence of the minimizing potential~\cite{Burke:2009} and result in a higher reliability for the topological loop to prevent damaging the surface or the tip \cite{Garrett:2016}. Recently, the introduction of heterodyne detection methods in both AM~\cite{Sugawara:2012} and FM~\cite{Sugawara:2020} KPFM has been shown to yield improved resolution and speed. In \cannex{}, we intend to use heterodyne AM. In contrast to the literature, our cantilever is excited at its resonance frequency $\omega_{c0}$ electrostatically. This method is less prone to artifacts and offers increased resolution compared to both classical AM and FM KPFM. Moreover, heterodyne KPFM enables us to detect the contact potential difference simultaneously to \(\partial^2 C/\partial a^2\), where \(C\) is the capacitance and $a$ is the distance between tip and surface \cite{Ma:2013, Sugawara:2020, Miyazaki:2022}.  
In the potential domain we expect the resolution to be better than $0.1\,$mV \cite{Nonnenmacher:1991}. 
This setup can easily be adjusted to any homo- or heterodyne detection method. If necessary, we will diverge from the intended use of heterodyne AM-KPFM if other methods prove to lead to higher resolutions.\\

\subsection{Optical Detection System: Force and Distance measurements}
\label{sec:setup:omsystem}
A major problem in the proof of principle for \cannex{}~\cite{Sedmik:2018kqt,Sedmik:2020cfj} was the parasitic coupling of AC signals into the cavity. We have therefore replaced the capacitive detection system by a purely optical one to detect all relevant parameters. Electrical potentials between the plates are now defined by a single source driven by the active potential compensation circuit described in \secref{sec:setup:chargecomp}. 
\subsubsection{Force and Force Gradient Detection}
\label{sec:setup:omsystem:detection}
\begin{figure}[!hb]
    \centering
    \includegraphics{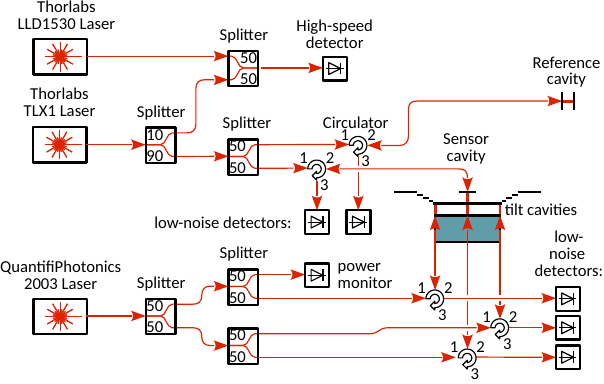}
    \caption{Complete schematic of the optical detection system.\label{fig:optical_scheme}}
\end{figure}

\cannex{} uses Fabry-P\'erot cavities formed by the polished ends of optical fibers and the reflecting surfaces of the sensor plate to measure the extension of the latter, and its distance to the opposing lower plate. An overview of the complete optical setup is given in \figref{fig:optical_scheme}.

A periodic force $F$ or movement of the sensor base $z_0$ at circular frequency $\omega$ leads to a displacement amplitude $\Delta z$ of the sensor plate according to the transfer functions 
\begin{align}
 T_{Fz}\equiv\frac{\Delta z}{F}=\frac{1}{m\left(\omega_0^2-\omega^2-\frac{\partial_aF}{m}-\ri\frac{\omega\omega_0}{Q}\right)}\,,\quad \text{and }T_{z_0z}\equiv\frac{\Delta z}{z_0}=\frac{\omega^2}{\left(\omega_0^2-\omega^2-\frac{\partial_aF}{m}-\ri\frac{\omega\omega_0}{Q}\right)}\,.\label{eq:sensor_tfs}
\end{align}
Note that at $\omega=0$, the transfer function reduces only approximately to Hooke's law, as $T_{Fz}\to(m\omega_0^2-\partial_a F)^{-1}=(k-\partial_a F)^{-1}$. We denote the sensor spring constant by $k$, the free resonance frequency by $\omega_0$, and the effective mass by $m$, which is larger than the physical plate mass $m_0$ due to the dynamical contribution of the spring elements. At the smallest separations $a\to\SI{3}{\micro\metre}$, the ratio $(\partial_a F)/k$ reaches values up to 0.01 such that the force gradient $\partial_a F$ cannot be neglected when evaluating DC extension data. We therefore have to either measure or calculate $\partial_a F$ for all measurements. The sensor resonance frequency $\omega_r$ (defined as the frequency at which the mechanical system has $\pi/2$ phase shift with respect to the sinusoidal force excitation signal) shifts according to 
\begin{align}
 \omega_r=\sqrt{\omega_0^2-\frac{\partial_aF}{m}+\inv{m}\mathcal{O}\big((\partial^2_a F)T_{Fz}F\big)}\,,
 \label{eq:frequency_shift}
\end{align}
where the relative error due to the last term is smaller by four orders of magnitude than the effect of the second at all separations.
We use the single interferometer above the sensor (see \figref{fig:optical_scheme}, `sensor cavity') to synchronously detect the DC extension $\Delta z$ in response to constant forces acting onto the plate, and the dynamical response $\Delta z(t)$ to an electrostatic excitation $F_\text{exc}(t)=(\varepsilon_0/4a^2)V_\text{exc}^2 \cos\omega_r t$ due to the voltage $V_\text{exc}$ applied between the plates at frequency $\omega_r/2$. Using a phase-locked loop (PLL), we can track $\omega_r$ and detect the shift $\Delta\omega=\omega_r-\omega_0$ of the resonance frequency, from which we extract $\partial_a F$ by inverting \eqnref{eq:frequency_shift}. For measurements in Xe gas, the sensor is over-critically damped, such that $\Delta \omega$ cannot be measured. In these measurements, we move to $a>\SI{10}{\micro\metre}$, where $\partial_a F/k\ll10^{-2}$ and the error by using computed values for the force gradients due to the dominant electrostatic and Casimir force contributions is negligible.

The value of $\Delta z(t)$ is extracted from the optical signal 
\begin{align}
 S=S_A+S_B\cos\frac{4\pi(d-\Delta z)}{\lambda}\,,
 \label{eq:opt_signal}
\end{align}
where the offset $S_A$ and amplitude $S_B$ are determined by the optical and geometric properties of the cavity, the laser power $P_L$ and wavelength $\lambda$, and the sensitivity of the detector. All appearing parameters are calibrated independently (see \secref{sec:setup:omsystem:calibration}). In order to maximize the sensitivity of $S$ to $\Delta z$ in \eqnref{eq:opt_signal}, we need to adjust $\lambda$ such that $\cos 4\pi d/\lambda=0$, which we call the `quadrature point'. Before data taking, we ensure the latter condition by performing a sweep of $\lambda$ at large distance, where $\Delta z$ can be calculated with sufficient precision. The sweep data are then fitted by \eqnref{eq:opt_signal}, with free parameters $S_A$, $S_B$, and $d$. Note that for all $a$ the interferometric cavity size $d$ only changes by $|\Delta z|\ll |d-\lambda/2|$ (the size of a fringe) due to the sensor's reaction to forces applied between the two plates. We can thus measure $d$ and adjust $\lambda$ to be at quadrature. The same sweep method with wide range of $\lambda$ allows us to measure the absolute distance $a_i$ between the two plates at the position of the lower three interferometers ($i=1..3$, see \figref{fig:optical_scheme} and \figref{fig:setup_overview}). We extract $a_i$ either from fits as described above or from the peaks appearing in the Fourier transformed data $\tilde{S}(d)$ of $S(\lambda)$. Which method is used depends on the cavity size. At large $a$, where many fringes can be covered by the modulation range (1520 -- 1620\,nm) of the QuantifiPhotonics 2003 laser, the Fourier method gives fast and accurate values of $a_i$ while at the smallest separations, not even one fringe can be covered and only the fit method can be applied.

The optical paths $2d$ and $2a_i$ of our cavities change with the vacuum pressure, and are significantly influenced during measurements in Xe gas. For this reason, we use an auxiliary fixed-distance cavity made of a material with effectively zero thermal expansion coefficient, sourced by the same laser driving the upper sensor cavity. Being located next to the sensor, this cavity gives a reference signal $S_\text{R}(P_L,\lambda,\rho_G)$ depending on the density $\rho_G$ of the gas, and fluctuations in both the laser power and wavelength.
For the three interferometers below the sensor, we use a simple power monitor to eliminate power fluctuations from the signal.
As all excitations and resonances are well below 20\,Hz, we use slow low-noise detectors with cutoff-frequency 1\,kHz to eliminate high-frequency noise.

Measurements in interfacial and Cavendish configuration are performed in sweeps starting at the maximum separation $a_\text{max}=\SI{30}{\micro\metre}$, reducing the separation for each measurement point in discrete logarithmic steps towards $a_\text{min}=\SI{3}{\micro\metre}$.
Before each sweep, a full re-calibration is performed (see below) to cancel drifts. The cavity size $d$ and the wavelength $\lambda$ are re-calibrated before each single measurement point. For measurements in Xe, $a$ is kept constant and sets of several consecutive measurements are performed at the same pressure. Each set is preceded by a full calibration with $d$ and $\lambda$ recalibrations in between single measurements
\subsubsection{Calibration}
\label{sec:setup:omsystem:calibration}
In order to perform an absolute measurement of forces, we need to calibrate all optical, mechanical and electric properties of our detection system. Some calibrations are invalidated only by ageing for which one measurement per experimental campaign is sufficient while others have to be repeated as often as possible to compensate drift. Constant offsets requiring only few re-calibrations concern the dependence of laser power on wavelength, transmission functions of wiring and electronics, etc. taken into account in the error calculations in \secref{sec:error}. The remainder of this section focuses on the frequent calibration of physical properties of the sensor and optical system. Mechanically, the sensor response is influenced by thermal fluctuations, long-term changes in the residual water layer on its surface, and surface charges. Even if these effects are expected to be very small, only a calibration can exclude them with certainty.

We start by re-calibrating the time-dependent wavelength offset of our lasers using a second laser with wavelength locked to an acetylene transition at $\lambda_\text{ref}=\SI{1532.83230(8)}{\nano\metre}$ and a beat technique~\cite{Dobosz:2017}. For this method, the output of the tested laser is combined with the one of the reference laser and lead to a high-frequency detector (see \figref{fig:optical_scheme}). Then, the $\lambda_\text{set}$ setting of the TLX1 is adjusted to result in a minimum beat frequency $\Delta f=(c/2)(\lambda_\text{set}^{-1}-\lambda_\text{ref}^{-1})$ using a lock-in amplifier. From the difference between $\lambda_\text{set}$ and $\lambda_\text{ref}$, and from the residual $\Delta f$ (resolution of $\lambda_\text{set}$), we can determine the (constant) error in $\lambda_\text{set}$ to within $\sim\SI{0.1}{\pico\metre}$.\\
Next, similar as in the proof of principle~\cite{Sedmik:2018kqt}, we increase the plate separation in high vacuum to $a_\text{cal}\approx\SI{5}{\milli\metre}$ where all interactions (electrostatic, Casimir, and gravity) between the plates fall off by at least two orders of magnitude with respect to their values at $a=\SI{30}{\micro\metre}$. In this position, the properties of the sensor cavity (see \figref{fig:optical_scheme}) and the reference cavity are determined by a wavelength sweep as described above. Subsequently, at $\lambda_\text{set}\approx\lambda_\text{ref}$, $d$ is adapted iteratively to match the one of the reference cavity such that both cavities are of the same size and at quadrature. Then, a well known electrostatic excitation is applied at frequency $\omega_\text{exc}$ that is swept over a range from $<\omega_0/2$ to $>2\omega_0$, and the signal amplitude and phase are decoded by a lock-in amplifier. Finally, a DC voltage is applied between the plates, and its value is swept over a range around zero, resulting in similar signal levels as in the actual measurement of the Casimir force. Both the extension $\Delta z$ and the frequency shift $\Delta \omega$ in response to the electrostatic force are recorded. Then, a synchronous fit of data from both sweeps (frequency and voltage) to \eqref{eq:sensor_tfs} and \eqref{eq:frequency_shift}, considering the signal non-linearity from \eqnref{eq:opt_signal}, all separately recorded voltages, power fluctuations (see below), and calculated forces and their gradients contributing to $\Delta z$ at $a=a_\text{cal}$, is performed. This fit results in accurate values for $m$, $\omega_0$, and $Q$.

\section{Error Budget}
\label{sec:error}
We already published a complete error budget~\cite{Sedmik:2021iaw} on the basis of a preliminary design considering a two-stage SAS, but a location inside a (seismically and thermally) noisy lab in Vienna. The leading reason to relocate \cannex{} to COBS are significantly lower environmental disturbances. Here, we update the previous error budget firstly with respect to the new location, and secondly for the final design and the characteristics of the actually used devices. As we show subsequently, the final design implements major improvements with respect to the previous conservative estimations, leading to an expected final sensitivities of {$\SI{0.259}{\nano\newton/\metre^2}$ and $\SI{0.0179}{\milli\newton/\metre^3}$ with projected uncertainties $\SI{0.119}{\nano\newton/\metre^2}\text{(stat.+sys.)}+\SI{0.139}{\nano\newton/\metre^2}\text{(const.)}$ and $\SI{8.6}{\micro\newton/\metre^3}\text{(stat.+sys.)}+\SI{9.3}{\micro\newton/\metre^3}\text{(const.)}$}, respectively at 68\% confidence level ($1\sigma$) for 100 days of measurements at $a=\SI{20}{\micro\metre}$. These figures represent an improvement by factors 2 and 30, for measurements of the pressure and pressure gradient between flat parallel plates, respectively, in comparison to our previous estimate.

\subsection{Seismic noise}
\label{sec:error:seismic}
As shown in \figref{fig:seismic_background} above, the seismic noise at the COBS is significantly lower than in Vienna at all frequencies above the micro-seismic peak. Due to a near seismically active zone in the Pannonian basin south of Vienna, earthquakes of low magnitude are frequent. However, the observatory houses official geomagnetic, seismic, and meteorological surveillance stations, including several Streckeisen STS-2 seismometers, which can be used not only to veto affected data but also to correct variations in the gravitational acceleration $g$ due to Earth tides and irregular local mass shifts at the location of \cannex{}.
\begin{figure}[!ht]
    \centering
    \includegraphics[width=\textwidth]{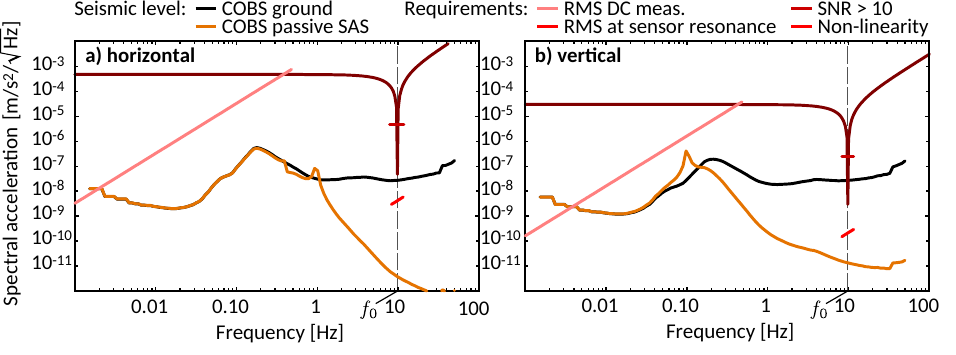}
    \caption{Numerical result of the passive \textbf{a)} horizontal and \textbf{b)} vertical seismic background on the core chamber, compared to updated requirements (red lines) representing upper limits (for details see~\cite{Sedmik:2021iaw}). The vertical dashed lines indicate the position of the vertical sensor resonance.\label{fig:seismic_performance}}
\end{figure}

The lower seismic noise at the COBS relaxes our requirements for the SAS such that the one-staged system described in \secref{sec:setup:sas} suffices. In \figref{fig:seismic_performance} we show the expected spectral seismic disturbance together with the limits from RMS noise, signal non-linearity, and signal-to-noise ratio (SNR). At all frequencies at COBS, the passive SAS alone already fulfills the requirement with 53\,dB and 22\,dB (amplitude) buffer in the horizontal and vertical directions, respectively, around the sensor resonance. Near the resonance of the GAS filter (assumed \SI{100}{\milli\hertz}), in the vertical direction, the buffer reduces to 14\,dB. Such a low resonance frequency is usually not achieved by a passive system but since active feedback can lower the resonance frequency and the corresponding amplitude even further, we use this assumption in our calculations. At frequencies below \SI{30}{\milli\hertz}, RMS noise becomes an issue. However, all data below \SI{10}{\second^{-1}} will be corrected by STS-2 data, which eliminates the constraint. Furthermore, the current error budget does not include additional damping by active feedback, as the respective design is not yet complete. We conservatively expect 2\,dB additional damping around the sensor resonance and about twice this reduction for the amplitude of the primary GAS filter and pendulum resonances.

\subsection{Detection noise}
\label{sec:error:detection}
For the error budget, we have to consider the time-dependent offsets and noise in all calibrated quantities, parameters, inputs, excitation and detection signal paths, and measurement devices. With respect to the previous error budget, we now have detailed and specific information about most quantities available, which allows us to compute the final error level expected for the measurements.

General boundary conditions are a total number of 100 distance sweeps, each including a re-calibration of cavity sizes, laser wavelength and sensor parameters, and elimination of seismic disturbances at frequencies lower than $(\SI{10}{\second})^{-1}$ using seismometer data. In the following, we construct the error budget by first analyzing the signal paths for DC and AC measurements independently, leading to a voltage and frequency signal, respectively. In the second step, the signals are then converted to a pressure and a pressure gradient using the calibrated mechanical properties of the sensor. In each of these two stages and for the calibration, we perform a detailed and -- to the best of our knowledge -- complete error analysis assuming small and normally distributed statistical errors and time-dependent drifts wherever manufacturer data are available. Details on each considered error are given in \appref{app:error_details}.

For DC signals of the extension $\Delta z$ we consider variations $\delta d$ in the cavity size $d$ due to seismic vibrations and thermal drift. Note that errors $\sigma d$ in the cavity size determination are constant offsets that need to be considered only for the conversion to a force below, as they are nullified for the voltage signal at quadrature. Wavelength errors $\delta \lambda$ are due to the laser bandwidth and spectral frequency noise as well as time-dependent wavelength accuracy errors $\sigma \lambda(t)$. The latter can be reduced by first assuring the independence~\footnote{Specifications obtained directly from the manufacturer.} of the offset $\sigma \lambda (\lambda)$ from the wavelength, for all $\lambda$ in the tuning range of the TLX1 laser, using a spectrometer with $<0.1\,$pm accuracy. Then, during operation, $\sigma \lambda$ will be measured repeatedly (before each force measurement) at one single wavelength using a frequency-locked reference laser as described above. According to manufacturer information, the uncertainty of the reference laser wavelength is mainly limited by thermal drift. The given uncertainty value $\sigma\lambda=0.08\,$pm is specified for $\delta T=\SI{3.5}{\degree}$, for which we reduce this value by a factor $0.1/3.5$ for operation at COBS. For the signal error, we further consider relative power fluctuations $\delta P_\text{L}$ of the laser leading to intensity fluctuations affecting both the signal of the measuring cavity and the one of the reference cavity $S_\text{R}$. As both signals are measured independently, this allows us to normalize the signal in realtime. Both optical signals contain stochastic noise $\delta V_\text{Det}$ of the detectors and $\delta V_\text{DAQ}$ of the two Keysight 34470A voltmeters, discretization error $\delta V_\text{alias}$, voltage offsets $\sigma V_\text{DAQ}(T)$ depending on fluctuations of the ambient temperature $T$, and a stochastic component for cable pickup noise. This leads to an expression for the total measured signal $V_\text{Sig}$, where for brevity, we combine for some quantities the respective constant and systematic errors $\sigma X(t)$ that may contain drift depending on time $t$, and stochastic noise $\delta X$ as $\delta X(t)$:
\begin{align}
V_\text{Sig}&=\frac{V_\text{R}(0)+\delta V_{\text{R}0}}{V_\text{R}(t)+\delta V_\text{R}}\left(\delta V_{\text{DAQ}}(t)+\delta V_{\text{Det}}+[1+\delta P_\text{L}(t)]\left[S_A+S_B \frac{4\pi [d+\delta d(t)]}{\lambda+\delta\lambda(t)}\right]\right)\,,\label{eq:det_error}\\
 \text{with }&\ V_\text{R}(t)=\delta V_{\text{DAQ}}(t)+\delta V_{\text{Det}}(t)+[1+\delta P_\text{L}]\left[S_{A,\text{R}}+S_{B,\text{R}} \frac{4\pi [d_\text{R}+\sigma_\text{R}]}{\lambda+\delta\lambda(t)}\right]\,,\nonumber
\end{align}
The reference signal $V_\text{R}(0)$ is measured before the start of the measurements with long integration time $\tau_R$ (see below), determined from the minimum between noise averaging and the rising influence of long-term variations (drift). 
The measured ambient temperature in the tunnel at COBS normally changes over periods of weeks rather than hours and has a typical fluctuation amplitude of \SI{5e-3}{\celsius} per day. As a worst case (in the case of work being performed in the tunnel), we consider a sinusoidal diurnal temperature deviation with peak amplitude \SI{0.1}{\degree} with zero transition at the start of the measurement. For the resulting offset errors, we use manufacturer specifications for the voltmeters amended by noise measurements with the actual devices. The lasers are temperature-stabilized but nonetheless are affected by changes in $T$. For the TLX1, the power noise has a $1/f$ characteristic for frequencies below 10\,Hz extrapolated from -105\,dBc at 10\,Hz to smaller frequencies, and flattening off at $-40\,$dBc at around $2\times10^5\,$s due to the power regulation circuit~\footnote{Private communication with the manufacturer. The specifications were confirmed by actual measurements over 3 hours. We computed the Allan deviation from these data, showing no clear minimum but flattening around 5\,ks.}. Laser frequency noise $\delta f_\text{L}$ is most pronounced around 1\,Hz and reduced for smaller frequencies by a dither keeping the DC value of the wavelength constant within 10\,kHz ($\sim\SI{0.1}{\femto\metre}$). Due to the periodic laser frequency offset calibration between measurements, continuous power normalization, internal temperature calibration of the data acquisition system, and all-year temperature stability at COBS, any long-term drift is expected to be insignificant for the period of data taking (100 days). We thus cut off drift contributions ($\delta V_\text{DAQ}(t)$, $\delta f_L(t)$, $\delta P_L$) at $\tau_i=\SI{1e-5}{\second}$ and integrate over the fluctuation spectra from $f=t^{-1}$ to $\infty$ with a cutoff function $f_I(f)=1/(1+2\pi f \tau)$ provided by the detector with $\tau=\tau_\text{Det}=10^{-3}\,$s and $\tau=t$ for a variable integration time (2\,s for DC measurements and 83\,s for AC measurements, 1000\,s with averaging for one measurement point) to get the respective RMS error. This procedure replaces the $1/\sqrt{\tau}$ factor considered normally for stochastic quantities, for all errors for which we have spectral information. 
The cavity size $d$ has an uncertainty due to seismic disturbances ($\delta d=\SI{4}{\pico\metre}_\text{RMS}$ for $\tau=2\,$s and $\delta d=\SI{2.8e-2}{\pico\metre}$ for $\tau=1000\,$s). Another contribution to $\delta d$ comes from thermal drift. Based on the thermal expansion coefficients and geometry, we expect an effective coefficient of $\sim \SI{5e-8}{\metre / \celsius}$ (with rather large uncertainty) which translates to \SI{5}{\pico\metre} maximum amplitude. This error needs to be evaluated carefully but our results here indicate, that in order to keep the effect of this error small, we need to re-calibrate $d$ after each measurement point. Note that for $\delta V_\text{Sig}$, constant errors in $d$ are irrelevant, as they cancel out by subtracting the signal from the one at $a_\text{cal}$.\\*
Without error normalization [i.e. by setting $V_\text{R}(t)=V_\text{R}(0)$], we obtain for an integration time of $\tau_i=1000\,$s a total (statistical and systematic) detection error $\delta V_\text{Sig}=4.85\times 10^{-6}\,$V, which is dominated by stochastic $\delta P_L$ at short times and $\delta \lambda$ drift at times larger than $\sim 100\,$s. Including the reference measurement, which has a fixed-length cavity without distance fluctuations or temperature drift, this figure can be reduced (assuming even 10\% mismatch between sensor and reference cavity) to $8.90\times 10^{-7}\,$V, which is dominated by seismic vibration at short times, and $\delta V_\text{DAQ}$ at $\tau\gtrsim 500\,$s. Error contributions for a single measurement and 500 sequential measurements comprising one force (gradient) measurement are given in \tabref{tab:dc-error}. For comparison, a pressure of $1\,{\rm nN}/{\rm m}^2$ would result in a signal of $3.03\times 10^{-6}\,$V. Note that $\delta V_\text{DC}$ has no significant dependence on $a$.
\begin{table}[!ht]
\caption{Components of the DC signal error for fixed $a=\SI{3}{\micro\metre}$ and $\tau_\text{DC}=2\,$s for a single datum ($N=1$) and for $N=500$ ($\tau=1000\,$s) representing one single measurement point, considering drift models and constant deviations $\sigma_\text{R}=d+10\,$nm, $S_{A,\text{R}}=1.1S_A$, and $S_{B,\text{R}}=1.1S_B$.\label{tab:dc-error}.}
\centering
\begin{tabular}{l c c c c}
\toprule
\textbf{Error} & \textbf{Symbol} & \multicolumn{2}{c}{\textbf{Value [V]}}  & \textbf{Type}\\
& & $N=1$ & $N=500$ &\\
\midrule
Detector noise & $\delta V_\text{Det}$ & $6.0\times10^{-8}$ & $2.7\times 10^{-9}$ & stat.\\
DAQ input noise & $\delta V_\text{DAQ}$ & $8.9\times 10^{-8}$ & $4.0\times 10^{-9}$ & stat.\\
Laser power fluct. (canceled) & $\delta P_\text{L}$ & 0 & 0 & stat.\\
Laser bandwidth & $\delta \lambda$ & $2.1\times 10^{-12}$ & $9.3\times 10^{-14}$ & stat.\\
Laser frequency noise & $\delta \lambda$ & $1.6\times 10^{-10}$ & $7.6\times 10^{-12}$ & stat.\\
Seismic vibrations & $\delta d$ & $3.3\times 10^{-5}$ & $1.3\times 10^{-7}$ & stat\\
Tot. ref. measurement noise (72\,h)& $\delta V_\text{R}$ & \multicolumn{2}{c}{$8.2\times10^{-8}$} & stat\\
\midrule
DAQ input error & $\sigma V_\text{DAQ}(t)$ & $2.3\times 10^{-11}$ & $1.3\times 10^{-11}$ & sys.\\
Laser wavelength drift & $\sigma \lambda(t)$ & $7.0\times 10^{-13}$ & $1.3\times 10^{-10}$ & sys.\\
Cavity size drift & $\sigma d(t)$ & $2.9\times 10^{-9}$ & $8.3\times 10^{-7}$ & sys.\\
Tot. ref. measurement error (72\,h) & $\sigma V_\text{R}$ & \multicolumn{2}{c}{$7.9\times 10^{-7}$} & sys. \\
\midrule
DAQ calibration & $\sigma V_\text{DAQ}$ & \multicolumn{2}{c}{$1\times 10^{-7}$} & const.\\
Tot. ref. measurement error (72\,h) & $\sigma V_\text{R}$ &\multicolumn{2}{c}{$1\times 10^{-7}$} & const.\\
\bottomrule
\end{tabular}
\end{table}
For AC measurements of the frequency shift $\Delta \omega=2\pi\Delta f$, we consider the inherent phase stability of the lock-in amplifier and PLL feedback circuitry $\delta f_\text{LI}$ and $\delta f_\text{PID}$, respectively. There is no simple expression, such as \eqnref{eq:det_error} that could be used for direct error propagation, since the frequency measurement involves numerical operation of the PLL. Therefore, we measured the the noise and stability of the actual lock-in amplifier and feedback using a first-order passive RC-lowpass filter as test device. This measurement results in higher noise than in measurements with the \cannex{} sensor, as the $Q$-factor is significantly lower. Aiming to give a (very) conservative estimate, we consider these measurements representative, nonetheless. Furthermore, we consider the uncertainty in the $\omega_0$ calibration obtained from simulations (see below). 
Constant offsets $\sigma f$ of the lock-in amplifier clock are reduced to $<5\times 10^{-10}\,$Hz by referencing the PLL to an external Rubidium clock. In addition, the Allan deviation of the clock between calibrations (once per 24\, h) could give an error at the level $0.05\,{\rm ppm}/^\circ{\rm C}$, which we take into account. Voltage noise sources as described for DC measurements, vibrations and laser frequency noise are considered indirectly by expressing amplitude noise of the sensor signal in terms of a phase $\phi$ at the zero transition\label{page:frequency_noise_from_voltage}, $\delta\phi = (\partial V/\partial \omega t)^{-1}\delta V$ and $\delta \phi=\left[\partial \text{Arg}(T_{Fz})(\omega)/\partial\omega\right]\delta\omega$, where Arg() is the argument function. This computation over-estimates the real phase error by at least a factor 2 but we consider it as worst-case. We obtain $\delta f_V=\omega_0\lambda/(16\pi Q \Delta z_\text{exc})\delta V$, with the excitation displacement amplitude $\Delta z_\text{exc}$ depending on $V_\text{exc}$ and $a$, and $\delta V_\text{Sig}=7.33\times10^{-7}\,,\ \sigma V_\text{Sig}=5.58\times 10^{-7}\,$V, evaluated as described above for DC measurements but with $\tau_i=83\,s$. Note that we adapt $V_\text{exc}(a)=V_\text{exc}(\SI{10}{\micro\metre})\times (a/\SI{10}{\micro\metre})^{3/2}$ to render the excitation and associated shift in the sensor resonance frequency independent of $a$.
The same uncertainties lowered by longer integration time are used for the calibration of $\omega_0$ (see \secref{sec:setup:omsystem:calibration}). 
We then obtain the total frequency shift measurement error by adding all constant, systematic, and statistical errors listed in \tabref{tab:ac-error} as described in \secref{sec:budget}, leading at the shortest separation $a=\SI{3}{\micro\metre}$ to a single point ($\tau=1000\,$s) frequency determination error $\delta f=4.68\times 10^{-7}\,$Hz dominated by $\delta f_\text{LI}$ at all integration times (up to $\tau\sim 10^4\,$s) and $\sigma f=8.31\times 10^{-9}\,$Hz. This has to be compared by a minimum signal $\Delta f=4.95\times 10^{-6}\,$Hz for a pressure gradient of $1\,{\rm mN}/{\rm m}^3$. 
\begin{table}[!h]
\caption{Components of the AC signal error for fixed $a=\SI{3}{\micro\metre}$ and $\tau_\text{AC}=83\,$s for a single datum ($N=1$) and for $N=12$ ($\tau=1000\,$s) representing one single measurement point, considering drift models.\label{tab:ac-error}.}
\centering
\begin{tabular}{l c c c c}
\toprule
\textbf{Error} & \textbf{Symbol} & \multicolumn{2}{c}{\textbf{Value [Hz]}}  & \textbf{Type}\\
& & $N=1$ & $N=12$ &\\
\midrule
Signal noise & $\delta f_V$ & $3.9\times 10^{-9}$ & $1.4\times 10^{-10}$ & stat.\\
$f$-detection & $\delta f_\text{PID}$ & $2.2\times 10^{-6}$ & $6.3\times 10^{-7}$& stat.\\
PLL frequency noise & $\delta f_\text{LI}$ & $1.8\times 10^{-9}$ & $5.2\times 10^{-10}$& stat.\\
\midrule
Signal drift & $\sigma f\delta_V(t)$ & $3.0\times 10^{-10}$ & $1.6\times 10^{-11}$ & sys.\\
PLL phase stability & $\sigma f_\text{LI}(\tau)$ & $8.8\times 10^{-10}$ & $1.1\times 10^{-8}$& sys.\\
Resonance freq. unc. & $\sigma \omega_0$ & $2.3\times 10^{-10}$ & $2.3\times 10^{-10}$ & sys.\\
\midrule
Signal noise & $\sigma f_V$ & \multicolumn{2}{c}{$4.5\times 10^{-9}$} & const.\\
PLL phase error & $\sigma f_\text{LI}$ & \multicolumn{2}{c}{$5\times 10^{-10}$}& const.\\
Resonance freq. error & $\sigma \omega_0$ & \multicolumn{2}{c}{$8.8\times 10^{-11}$} & const.\\
\bottomrule
\end{tabular}
\end{table}

The measured frequency shift can be converted to a total force gradient by inverting \eqnref{eq:frequency_shift}, where we require the effective mass $m$, and $\omega_0$ from the calibration. In order to determine the error on $m$, $Q$ and $\omega_0$, we have performed a Monte Carlo simulation of complete calibration data on $V_\text{Sig}$ and $\omega_r$ considering all voltage and frequency measurement errors discussed in this section as normally distributed random quantities with the known width, and offsets depending on time. For frequency data, we created voltage signals containing $\delta V_\text{sig}$ and extracted the resulting amplitude and phase using a software lock-in amplifier. We then selected 300 arbitrary sets of frequency shift and voltage shift data, from which we extracted values for $m$, $Q$, $\omega_0$, as described in \secref{sec:setup:omsystem:calibration}. Eventually, we computed the standard deviation of the fit results, which we interpret as a systematic error (i.e. statistical, averaging with the number of calibrations only). The difference between the mean fit value and the originally used parameter value is representative of a constant error for this parameter. We obtain $\delta m=58.6\times 10^{-12}\,$kg, $\sigma m=1.42\times 10^{-12}\,$kg, $\delta Q=1.30\times 10^{-5}$, $\sigma Q=2.83\times10^{-2}$, $\delta \omega_0=1.44\times 10^{-9}\,{\rm rad}/{\rm s}$, $\sigma \omega_0=4.4\times 10^{-11}\,{\rm rad}/{\rm s}$. For $m$ and $\omega_0$, the constant errors are significantly smaller than the systematic ones, which indicates that repeated calibrations may be required to average out the systematic errors. For further computation, we use the constant frequency detection error $\sigma f=5.53\times10^{-10}\,$Hz for $\sigma \omega_0/(2\pi)$ instead of the smaller constant error from the simulation given above. We consider a measurement scheme, in which one calibration is performed per distance sweep (i.e. per day) and assume that due to the thermal stability of the system, $\sigma m$ and $\sigma \omega_0$ can be reduced as $1/\sqrt{N_\text{cal}}$ with the number $N_\text{cal}$ of calibrations. We then resolve $\omega_r+\delta \omega_r(t)=\sqrt{(\omega_0+\sigma\omega_0)^2-\partial_aF/(m+\sigma m)}$ for the total gradient $\partial_a F$, and propagate all errors. To evaluate the latter expression, we require a value for $\omega_r$, where we assume the Casimir force gradient (see \secref{sec:casimir}) and an electrostatic interaction $\partial_a F_\text{ES}=\varepsilon_0 A V^2/a^3$ with $V=\SI{0.5}{\milli\volt}\times a^2/(\SI{10}{\micro\metre})^2$ for the excitation. Using the quantities above and the frequency determination error $\delta \Delta f(t)$, we finally obtain the errors listed in \tabref{tab:df-error} for $a=\SI{3}{\micro\metre}$, yielding a total pressure gradient detection errors 
$\delta \partial_a F/A =0.097\,{\rm mN}/{\rm m}^3$(stat.+sys.) and $\sigma \partial_a F/A=0.001\,{\rm mN}/{\rm m}^3$(const.), dominated clearly by the frequency measurement error $\delta f_\text{PID}$, which is based on our test measurements at low $Q$-factor.
\begin{table}[!hb]
\caption{Components of the pressure gradient error for fixed $a=\SI{3}{\micro\metre}$ and $\tau_\text{AC}=83\,$s for a single datum ($N=1$) and for $N=12$ ($\tau=1000\,$s) representing one single measurement point, considering drift models.\label{tab:df-error}.}
\centering
\begin{tabular}{l c c c c}
\toprule
\textbf{Error} & \textbf{Symbol} & \multicolumn{2}{c}{\textbf{Value [N/m$^3$]}} & \textbf{Type}\\
& & $N=1$ & $N=12$ &\\
\midrule
Frequency det. error & $\delta f$ & $3.3\times 10^{-4}$& $9.4\times 10^{-5}$&stat.\\
\midrule
Mass cal. uncertainty & $\sigma m$ & $4.8\times 10^{-5}$& $4.8\times 10^{-5}$& sys.\\
Resonance freq. uncertainty & $\sigma \omega_0$ & $4.6\times 10^{-8}$& $4.6\times 10^{-8}$& sys.\\
Frequency det. error & $\sigma f(t)$ & $1.4\times 10^{-7}$& $1.6\times 10^{-6}$&sys.\\
\midrule
Mass cal. error & $\sigma m$ & \multicolumn{2}{c}{$1.2\times 10^{-6}$}& const.\\
Resonance freq. error & $\sigma \omega_0$ & \multicolumn{2}{c}{$1.1\times 10^{-7}$}& const.\\
Frequency det. error & $\sigma f$ & \multicolumn{2}{c}{$1.1\times 10^{-7}$} &const.\\
\bottomrule
\end{tabular}
\end{table}

\begin{figure}[!hb]
    \centering
    \includegraphics[width=\textwidth]{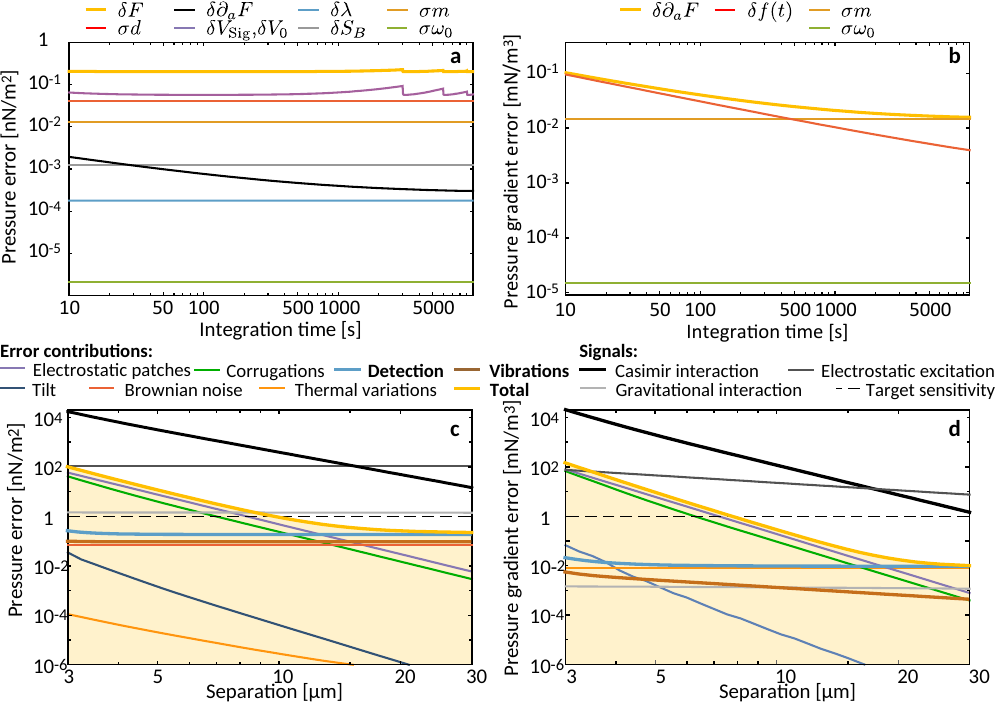}
    \caption{Updated error budget for \cannex{}. \textbf{a)} and \textbf{b)} show the dependence of the detection error in the pressure and pressure gradient, respectively for a single calibrated measurement at fixed separation $a=\SI{3}{\micro\metre}$. \textbf{c)} and \textbf{d)} reproduces error contributions from Ref.~\cite{Sedmik:2021iaw} with only detection errors and seismic errors updated. Note that here, $\delta X$ denotes the total error of quantity $X$ and $\sigma X$ the combined systematic and constant error.\label{fig:error_time}}
\end{figure}

The measured DC signal voltage can be interpreted as an extension $\Delta d=F \left.T_{Fz}\right|_{\omega\to0}$ of the sensor due to a force $F$, and its transfer function $T_{Fz}$, given in \eqnref{eq:sensor_tfs}. In order to evaluate the latter quantity, we require the errors of $\partial_a F$, $m$, $d$, as well as the optical amplitude $S_B$ and $V_\text{Sig}$. 
Eventually, we propagate the errors according to
\begin{align}
F(a)=\inv{4\pi}\left[\partial_a F\!+\!\delta \partial_aF(t) - (m\!+\! \sigma m)(\omega_0\!+\!\sigma\omega_0)^2\right]\left[4\pi\sigma d-(\lambda\!+\!\delta\lambda(t))\text{asin}\left(\frac{V_\text{Sig}\!+\!\delta V_\text{Sig}(t)}{S_B+\sigma S_B}+\sin\frac{4\pi\sigma d}{\lambda+\delta\lambda(t)}\right)\right],\label{eq:error_force}
\end{align}
resulting in an error of $\delta F/A =0.324\,{\rm nN}/{\rm m}^2$ and $\sigma \partial F/A=0.167\,{\rm nN}/{\rm m}^2$, for a single measurement of 1000\,s at $a=\SI{3}{\micro\metre}$ dominated by the uncertainty $\delta V_{Sig}$ and uncertainty $\sigma d$ in the cavity size. The value of $\sigma d$ is the parameter error obtained by the wavelength sweep fit during repeated calibrations, for which we categorize it as systematic error, influenced by thermal variation. The corresponding offset (constant error) was below machine precision in the fit. We thus consider in the final budget a factor $1/\sqrt{N_\text{sweep}}$ with $N_\text{sweep}=500$ for $\sigma d$. All errors contributing to $\delta F/A$ are listed in \tabref{tab:f-error}. These results highlight again the need for thermal stability, and vibration attenuation as well as proper thermal design. In \figref{fig:error_time}a and b we show the detection errors for the pressure and its gradient as a function of integration time. In light of the considerations above, the previous error budget is thereby improved by roughly one order of magnitude in both measured quantities. For the pressure error, the most important contributions are the drift and uncertainty ($\sigma d$) in the cavity size, DC measurement noise which in turn depends on vibrations, and uncertainty $\sigma m$ in the mass of the sensor. For the pressure gradient, the most important contributions come from the frequency measurement that depends strongly on the internal stability of the PLL and the sensor mass. Temperature drift is a crucial parameter influencing most systematic errors considered here.
\begin{table}[!h]
\caption{Components of the pressure error for fixed $a=\SI{3}{\micro\metre}$, $\tau_\text{DC}=2\,$s and $\tau_\text{AC}=83\,$s for a single datum ($N=1$) and for $N=12$ ($\tau=1000\,$s) representing one single measurement point, considering drift models.\label{tab:f-error}.}
\centering
\begin{tabular}{l c c c c}
\toprule
\textbf{Error} & \textbf{Symbol} & \multicolumn{2}{c}{\textbf{Value [N/m$^2$]}}  & \textbf{Type}\\
& & $N=1$ & $N=12$ &\\
Force gradient error & $\delta \partial_a F$ & $6.0\times 10^{-12}$& $5.0\times 10^{-13}$& stat.\\
DC signal error & $\delta V_\text{Sig}$ & $1.8\times 10^{-10}$& $2.0\times 10^{-12}$& stat.\\
Zero force DC signal error & $\delta V_0$ & $1.8\times 10^{-10}$& $2.4\times 10^{-11}$& stat.\\
\midrule
Mass cal. uncertainty & $\sigma m$ & $4.1\times 10^{-11}$& $1.2\times 10^{-11}$& sys.\\
Resonance freq. uncertainty & $\sigma \omega_0(t)$ & $8.6\times 10^{-16}$& $2.5\times 10^{-16}$& sys.\\
Cavity size error& $\sigma d(t)$ & $4.2\times 10^{-10}$& $1.2\times 10^{-10}$& sys.\\
Wavelength drift & $\sigma \lambda(t)$ & $1.4\times 10^{-12}$& $4.0\times 10^{-13}$& sys.\\
Fringe amplitude uncertainty & $\sigma S_B$ & $1.3\times 10^{-11}$& $3.6\times 10^{-12}$& sys.\\
Force gradient error & $\sigma \partial_a F$ & $6.6\times 10^{-13}$& $1.9\times 10^{-13}$& sys.\\
DC signal error & $\sigma V_\text{Sig}$ & $9.0\times 10^{-11}$& $4.9\times 10^{-11}$& sys.\\
Zero force DC signal error & $\sigma V_0$ & $9.1\times 10^{-11}$& $1.7\times 10^{-10}$& sys.\\
\midrule
Mass cal. uncertainty & $\sigma m$ & \multicolumn{2}{c}{$9.0\times 10^{-12}$}& const.\\
Resonance freq. uncertainty & $\sigma \omega_0(t)$ & \multicolumn{2}{c}{$2.1\times 10^{-15}$}& const.\\
Cavity size error& $\sigma d(t)$ & \multicolumn{2}{c}{$6.7\times10^{-11}$}& const.\\
Wavelength accuracy & $\sigma \lambda(t)$ & \multicolumn{2}{c}{$3.9\times 10^{-14}$}& const.\\
Force gradient error & $\sigma \partial_a F$ & \multicolumn{2}{c}{$2.2\times 10^{-13}$}& const.\\
DC signal error & $\sigma V_\text{Sig}$ & \multicolumn{2}{c}{$4.8\times 10^{-11}$}& const.\\
Zero force DC signal error & $\sigma V_0$ & \multicolumn{2}{c}{$4.8\times 10^{-11}$}& const.\\
\bottomrule
\end{tabular}
\end{table}

\subsection{Updated error budget}
\label{sec:budget}
Apart from seismic and detection errors updated above, we also improved our statistical methods. Following~\cite{Rabinovich:2005}, we compute the total error $\sigma^\text{tot}_p$ at probability $p$ as
\begin{align}
\sigma^\text{tot}_p = \sum_i\sigma^\text{const}_i +\sqrt{t_p^2(\nu_\delta)\sum_j^{N_\delta}[\delta^\text{stat}_j]^2+t_p^2(\nu_\sigma)\sum_k^{N_\sigma} [\sigma^\text{sys}_k(\tau)]^2}\,,
\end{align}
where $\sigma^\text{const}_i$ are the independent constant (undetermined systematic) errors~\footnote{Note that constant errors occur for many devices and are not limited to the aliasing error, as mentioned in the literature~\cite{Bordag:2014}. For example, consider an internal calibration offset of a voltmeter due to aging that may change on time scales larger than the experimental period. Even if a traceable certified calibration is performed before the experiment, the error cannot be determined during the experiment but has to be considered as a maximum offset.}. Such errors are not statistically distributed (varying) over time scales of the experiment and can only be estimated conservatively from the accuracy limit given by the manufacturer. They are linearly added and do not reduce with time. $\delta_j$ are statistical random errors varying on time scales shorter than any integration time $\tau$, such that they properly probe a (normal) distribution and can be reduced by a factor $1/\sqrt{\tau}$. $\sigma^\text{sys}_k(t)$ are the (statistical components of) systematic errors. In this category, we have any offset that has changes that are quick enough to exhibit a distribution during the experiment that may probably not be sampled completely. For example, we have aliasing errors and temperature drift as well as errors of parameters determined in repeated calibrations. These errors average with the number of calibrations or the number of measurements obtained at the same conditions and parameters. $t_p(\nu)$ is the $p$-\% point of the student distribution that depends on the number $\nu$ of degrees of freedom, $\nu_x=N_x-1$. Note that for $p=0.68$ at $1\,\sigma$ level, $t_p<1$ for which the total error is smaller than the single errors in \figref{fig:error_time}. For all $\delta_i$ determined from $N_i$ individual measurements $x_j$, each with a total error $\sigma_j$, it is common to consider the (weighted) error of the mean
\begin{align}
    \delta=\left[\sum_{j=1}^{N_i}\frac{\sigma_j^{-2}(x_j-\bar{x}_w)}{\sum_{k=1}^{N_i}\sigma_k^{-2}}\right]^{\frac{1}{2}}\,,\quad\text{with }\bar{x}_w=\sum_{j=1}^{N_i}\frac{\sigma_j^{-2}x_j}{\sum_{k=1}^{N_i}\sigma_k^{-2}}\,,\label{eq:weighted_mean}
\end{align}
where we have introduced the weighted mean $\bar{x}_w$. While for experimental data points, weighted quantities can differ from unweighted ones due to singular noise events, in the present \emph{estimation}, $\sigma_j=\sigma\,\forall j$, for which \eqnref{eq:weighted_mean} reduces to the geometric mean and its error.

With seismic and detection errors updated, we achieve the prospective error budget in \figref{fig:error_time}c and d. We replot errors discussed in Ref.~\cite{Sedmik:2021iaw} with only the seismic and detection errors updated. All detection errors have a mild dependence on separation due to the adaptation of $V_\text{exc}$. {For details on other errors, we refer the reader to the detailed discussion in Ref.~\cite{Sedmik:2021iaw}}. The detection error is the main limitation at separations $a\gtrsim\SI{10}{\micro\metre}$, for which the updated error budget presented here improves the prospects for measurements at large separations. {Deformation errors, including sag of the surfaces due to the measured pressures and gravity are the second-strongest error contribution at small separations after residual patch effects. Using Talbot interferometry on the actual plate surfaces, we are able to measure the deformation and take it into account, for which the error given here (green line), which considers a \emph{residual} spherical deformation of $4\,$nm amplitude can be considered as an absolute worst case. Note further, that local differences in the Xe density in gas pressure modulation measurements (see below) near the surfaces due to temporal adsorption, rarefaction, or other stratification effects would cause only negligible errors not influencing the budget presented here.} As the present error budget is still partially based on models, we nonetheless present the updated prospects below considering the previous worse error budget. 
\section{Prospective Results}
Recently~\cite{Sedmik:2021iaw}, we gave prospective limits on axion-like interactions, Symmetron DE interactions, and measurements of the Casimir effect. Although the error budget in \secref{sec:error} demonstrates a further improvement in both pressure and pressure gradient measurements, we conservatively keep the baseline of \SI{1}{\nano\newton/\metre^2} \SI{1}{\nano\newton/\metre^2} and \SI{1}{\milli\newton/\metre^3}. In the following, we present updated calculations regarding equilibrium and non-equilibrium Casimir forces and limits on a range of DE interactions with updated theoretical methods and consider the final design. The latter limits supersede previous ones in Refs.~\cite{Almasi:2015zpa,Sedmik:2021iaw}.
\subsection{Casimir Effect}
\label{sec:casimir}
Casimir force experiments open a very extensive window into the quantum mechanical behavior of physical systems. Since the prediction of the Casimir effect in 1948~\cite{Casimir:1948} the theoretical framework characterizing this phenomenon has substantially evolved and nowadays it is situated at the intersection of very diverse areas of physics, ranging from material science and statistical physics to quantum field theory. An accurate measurement of the Casimir force has therefore the potential not only to offer more information about the behavior of the system's quantum fluctuations but also to test how different theories merge together, possibly providing a new window into fundamental physics. 

One of the most remarkable aspects of the Casimir interaction is its dependence on the involved materials, the thermodynamic state and the geometry of the system. Indeed, investigations have shown that by modifying these properties the Casimir force can be tuned, with interesting implications both for fundamental research and modern quantum technologies. Below we provide a short review of how these three aspects affect the Casimir interaction and the role that \cannex{} may play in approaching them separately or also simultaneously. 

\subsubsection{Material properties}

Already from Casimir's original article on the force between two parallel perfect reflecting plates, it appears clear that the properties of the materials involved in the system can play a role in determining the behavior of the force. The work of Lifshitz in 1955~\cite{Lifshitz56} underlined this aspect even further. The celebrated Lifshitz formula 
\begin{align}
P_{\rm Lif}(a,T)=-\mathrm{Im}\int_{0}^{\infty}\frac{d\omega}{\pi}\int\frac{d\mathbf{k}}{(2\pi)^{2}}\sum_{\sigma}\hbar\coth\left[\frac{\hbar\omega}{2k_{\rm B}T}\right]\;
\kappa\frac{r^{\sigma}_{1}(\omega,k)r^{\sigma}_{2}(\omega,k)e^{- 2\kappa a}}{1-r^{\sigma}_{1}(\omega,k)r^{\sigma}_{2}(\omega,k)e^{-2\kappa a}}~,
\label{Elifshitz}
\end{align}
provides the force per unit of area between two parallel planar structures separated by a distance $a$ in terms of the planes' reflection coefficients $r^{\sigma}_{i}(\omega,k)$. In the previous expressions $\sigma$ defines the polarization (TE or TM) of the electromagnetic field, $\mathbf{k}$ is the component of the wave vector parallel to the surfaces, $k=|\mathbf{k}|$ and $\kappa=\sqrt{k^{2}-\omega^{2}/c^{2}}$ ($\mathrm{Im}[\kappa]\le 0; \mathrm{Re}[\kappa] \ge 0$). Considering materials with different reflection properties, several experimental groups have shown that the Casimir pressure can be substantially modified~\cite{Iannuzzi04,deMan:2009zz,Munday:2009fgb,Torricelli:2010a,Lamoreaux10,Banishev12a,Banishev:2012bh,Banishev:2013,Somers18,Zhao19}. In particular, leveraging the interplay between optical properties and geometry (see also the next section) not only the strength but also the sign of interaction can be changed~\cite{Dzyaloshinskii61,Rodriguez13b,Esteso15,Zhao19,Esteso23}.

One of the most representative and, at the same time, most controversial examples highlighting the relevance of material properties in the Casimir interaction is provided by their role in determining the finite-temperature correction to the Casimir force between parallel metal plates. For more than two decades now this has been a topic of intense investigation and debate. A description of the metal in terms of a commonly used Drude model 
\begin{equation}
  \varepsilon(\omega) = 1 - \frac{\Omega^2 }{
  \omega(\omega + {\rm i}\gamma)},
  	\label{eq:drude}
\end{equation}
where $\Omega$ is the plasma frequency and $\gamma$ a nonzero dissipation rate, gives rise to a temperature dependence of the force
which considerably differs from that obtained for perfect reflectors~\cite{Bostrom00}. (For recent reviews on the debate around the thermal correction of the Casimir force, cf.\ Refs.~\cite{Brevik22,Klimchitskaya22c} and references therein.) This is particularly relevant at large temperatures and/or distances where the force predicted by the Drude model is half the value obtained for perfect reflectors. Puzzlingly, such behavior is not found in many precise measurements of the Casimir force~\cite{Decca07,Banishev12a,Banishev13,Bimonte:2016myg,Liu19a,Bimonte21}. Experiments where the reduction in strength predicted by the Drude model was observed~\cite{Lamoreaux10,Sushkov11,Garcia-Sanchez12} needed to consider systematic effects such as the presence of  patch potentials in their setups~\cite{Behunin:2012zr,Behunin14a,Behunin:2014tsa,Garrett15}. 
Perhaps even more surprising is that the experiments disagreeing with the prediction of the Drude model in~\eqnref{eq:drude} are in very good agreement with the result obtained by setting $\gamma=0$ in the same model. This quite suggestive behavior highlights the role of material properties and, for the present model, of dissipation in the controversy. 
More generally, within the Lifshitz framework, the disagreement between the experimental measurements and theoretical predictions obtained using the Drude model is related to the description of the optical response of metals at low frequency. This can substantially affect the contribution of the transverse electric ($\sigma=$TE) polarization~\cite{Torgerson04,Bimonte07a,Intravaia09,Klimchitskaya23f}.
More specifically, in the limit of large separations $a$, the difference between the two models discussed here arises because for the Drude model with $\gamma\neq0$, in agreement with the Bohr--van Leeuwen theorem~\cite{Leeuwen21,Bimonte09}, the contribution of the TE-polarization in~\eqnref{Elifshitz} vanishes at large distance~\cite{Klimchitskaya23f}. The model resulting by setting $\gamma=0$ in~\eqnref{eq:drude}, often called plasma model, is equivalent to a simple description of a superconductor~\cite{London35}, which does not fulfill the Bohr--van Leeuwen theorem. 

\begin{figure}[ht]
\centering
\includegraphics[width=0.45\textwidth]
{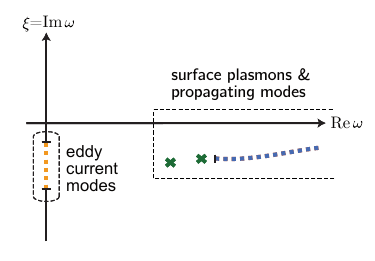}
\caption{Schematic representation of the typical electromagnetic mode-frequencies vibrating within a planar cavity made by dispersive and dissipative metallic mirrors described in terms of the Drude model~\cite{Intravaia22a}. Due to dissipation and according to causality, all modes are resonances described by a complex frequency located in the lower half of the complex-frequency plane. Typical resonances include surfaces plasmons (green crosses) and cavity modes (blue dots)~\cite{Intravaia05,Intravaia07}. Due to their diffusive nature, the eddy currents are described in terms of a pure imaginary frequency and are therefore located along the negative imaginary axis~\cite{Intravaia09,Intravaia10a}. \label{ModesPath}}
\end{figure} 

A complementary perspective can be given in terms of specific solutions of the Maxwell equations corresponding to purely dissipative (i.e., over-damped) 
modes~\cite{Intravaia22a} which are physically connected with Foucault current or `eddy current' in the interior of the material plate~\cite{Intravaia09,Henkel10,Intravaia10a,Reiche20} (see also Refs.~\cite{Bimonte07,Guerout14} for related investigations).
These modes have pure imaginary frequencies (see~\figref{ModesPath}) and their dynamics is described 
by a diffusion equation. The diffusion constant is given by $D=\gamma \lambda^{2}$, where $\gamma$ is the dissipation rate 
of the metal and $\lambda\equiv c/\Omega$ the plasma penetration depth. The electromagnetic field 
associated with these currents is evanescent in vacuum, i.e., it exponentially decays with the 
distance from the surface of the metal. 
In superconductors eddy current modes are suppressed by the Meissner effect, explaining the behavior of the Casimir effect with the plasma model.
It was shown that the eddy current contribution alone accounts for the difference in the prediction for the Casimir effect at finite temperature obtained with the Drude and the plasma model~\cite{Intravaia09,Intravaia10a}. In particular, in agreement with previous observations~\cite{Torgerson04,Bimonte07a}, the largest contribution of these modes arises for the TE-polarization~\cite{Klimchitskaya23f}. Eddy currents are also very useful to understand why accounting for spatial dispersion in light-matter interaction~\cite{Svetovoy05} can remove pathologies occurring in the thermodynamic behavior of the Casimir entropy when the Drude model is used~\cite{Bezerra04,Intravaia08,Klimchitskaya09b,Klimchitskaya09c,Milton11b}.
In particular, they allows to discern among the different models describing spatial dispersion, showing also that not all of them are able to eliminate these inconsistencies~\cite{Reiche20}.

Due to the accuracy and the flexibility of the measurements as well as the possibility to approach the system in its simplest geometry (two parallel plates) \cannex{} allows to approach the study of the interplay between material properties and the Casimir effect from a new perspective. The same flexibility also allows to probe the impact on the interaction of materials with special or exotic properties such as magnetic materials~\cite{Shelden23,Banishev12a,Banishev13c}, {graphene~\cite{Bordag09,Fialkovsky:2011pu,Egerland19,Liu:2021ice}} and others~\cite{Woods16,Fialkovsky18,Chen20a,Farias20a}, using simple planar structures.
This can offer new understanding for the resolution of the controversy and, in general, additional information about the behavior of the Casimir force in regimes and, in particular, for distances that were not explored before in experiments. 

\subsubsection{The geometry of the system}

It was recognized early on that the Casimir effect can be substantially modified by changing the geometry of the involved objects. One of the most remarkable examples is probably the calculation of T. Boyer in 1968 predicting a repulsive Casimir force on a perfectly conducting spherical shell cavity~\cite{Boyer68} (see also Ref.~\cite{Graham13} for recent evaluations with the same geometry).
In the last decade theoretical developments have shown how to efficiently compute the Casimir interaction in systems involving complex structures. A large variety of methods, ranging from semi-analytical~\cite{Buscher04,Davids10,Lambrecht08a,Intravaia12a,Messina15,Hartmann17,Antezza20,Schoger22,Emig23} to full numerical~\cite{Reid11,Johnson11,Rodriguez11,Reid13,Emig23,Kristensen23} have been developed. The main drives of this progress have been, on the one side, the necessity to accurately interpret the measurements of the Casimir force in realistic setups and, on the other side, the ambition to deterministically tune the interaction. Controlling the Casimir force can help in reducing unwanted stiction in microscopic devices like MEMS and NEMS and it can serve as an additional contactless mechanical actuator for similar devices~\cite{Chan01s,Chan01}.

Among the most studied geometries different from plane-plane originally considered by Casimir, one finds the plane-sphere configuration. In fact, this geometry has been for a long time the workhorse in experiments aiming to measure the Casimir force~\cite{Derjaguin56,Lamoreaux:1996wh,Mohideen:1998iz,Chan01s,Decca05,Bimonte21}. Considering a sphere in front of a plane releases indeed the constraint of parallelism, drastically simplifying the experimental setup. The price to pay is, however, a smaller signal and a more difficult interpretation of the measurement. The latter has for a long time relied on the so-called proximity force approximation, sometimes also called Derjaguin approximation~\cite{Derjaguin56}. If the radius of the sphere is larger than the distance between the surfaces of the two objects, this approximation connects the sphere-plane Casimir force to the energy in the plane-plane configuration. Although previous experiments have directly investigated the plane-plane configuration~\cite{Sparnaay57,Bressi:2002fr}, \cannex{} is one of the first modern apparatuses designed to reexamine this geometry without strongly penalizing compromises between control, accuracy, and strength of the signal. 
This same characteristic and the flexibility of this setup can be employed in order to investigate from a new perspective the interaction between different planar structures, ranging from multilayer stacks to nanostructured surfaces, like periodic gratings~\cite{Buscher04,Lambrecht08a,Chan08,Bao10,Lussange12,Intravaia12a,Intravaia13,Messina15,Antezza20,Wang21} or more modern and complex arrangements, as for example metasurfaces~\cite{Chen16c,Kort-Kamp21,Qiu21}.

\begin{figure}[ht]
\centering
\includegraphics[width=0.45\textwidth]
{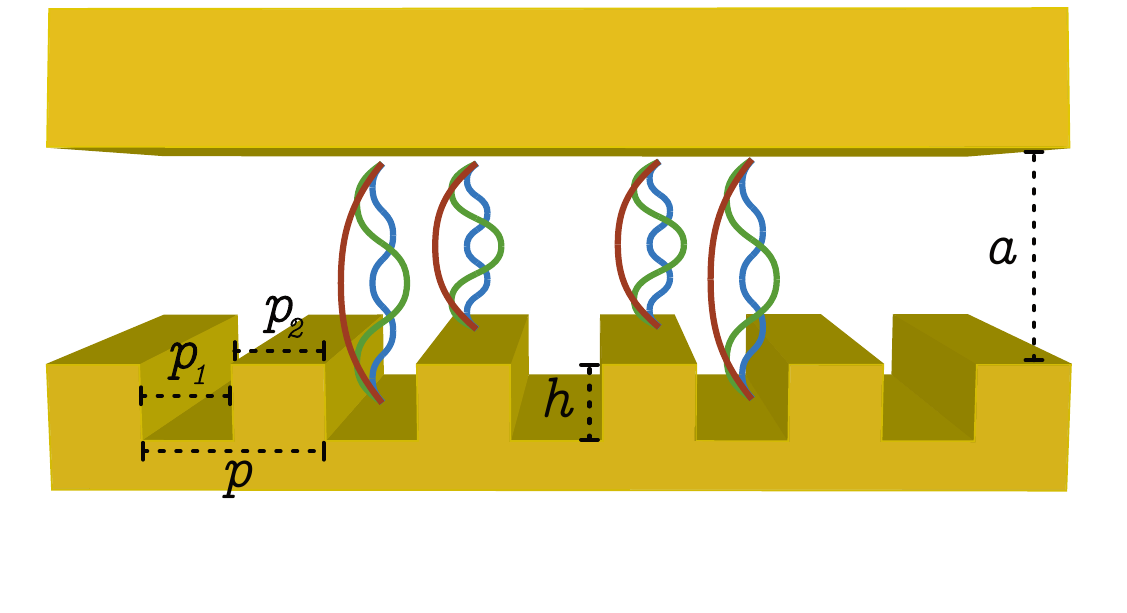}
\includegraphics[width=0.46\textwidth]
{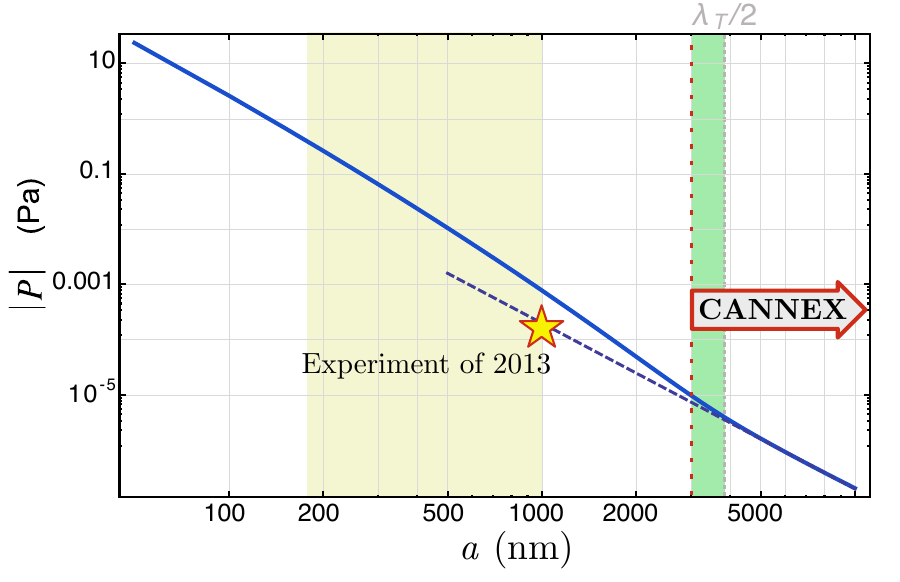}
\caption{
(Left) A schematic representation of one of the simplest configurations for investigating the impact of nanostructuring on the Casimir effect: A one dimensional lamellar grating facing a plane. The grating can be characterized with the help of the following parameters: width of the grooves $p_{1}$, width $p_{2}$ and height $h$ of the teeth.
(Right) Casimir pressure between a metallic grating and a metallic plane (see Ref.\cite{Intravaia12a} for further details).
The metal is modeled using the Drude model with $\Omega=8.39\,$eV and $\gamma=0.043\,$eV, corresponding to the values of gold. The grating is characterized by the following parameters: $p_{1}=160$ nm, $p_{2}=90\,$nm, and height $h=216$ nm. The temperature of the system is set to $T=300$ K. 
At separations larger than the thermal wavelength [$\lambda_{T}=\hbar c/ (k_{B}T) $] the pressure tends towards the value $\zeta(3)k_{B}T/(8\pi a^{3})$ (dashed curve), which is the same limiting behavior for the Casimir pressure predicted for the plane-plane configuration using the Drude model. At short separations, the pressure is $\propto a^{-3}$ because of the finite grating conductivity. The yellow shadow region describes the distance range investigated in Ref.~\cite{Intravaia13} and the star indicates the value of the Casimir pressure measured around \SI{1}{\micro\metre} in the same experiment (see the main text). 
 \label{gratingPressure}}
\end{figure}

Specifically, the simple one-dimensional lamellar grating structure has already found its way into Casimir physics. Its relative simplicity has allowed for an accurate theoretical description of the Casimir interaction between two vacuum-separated gratings with commensurable periods.
Within the framework of the scattering approach~\cite{Rahi:2009hm,Rahi11,Lambrecht11a,Ingold15a} the evaluation is essentially reduced to the calculation of the scattering matrices of the two nanostructured objects. For instance, the Casimir pressure at temperature $T$ between two parallel gratings with the same period $p$ separated by the distance $a$ can be obtained from~\cite{Lambrecht:2008eux,Davids10}
\begin{align}
P(a)&=-4k_{\rm B} T \sum^{\infty} _{l=0} {}^{'}
\int_{0}^{\infty}\hspace{-0.3cm}dk_y \int_{0}^{\pi/p} \hspace{-0.3cm}d\alpha_0  
\hspace{1mm}\partial_{a}\log {\rm det} \left[1- \underleftarrow{\mathcal{R}}^{L} \mathcal{P}(a)\,  \underrightarrow{\mathcal{R}}^{R} 
\mathcal{P}(a) \right] ~.
\label{pressure}
\end{align}
Here $\mathcal{P}$ are the matrices describing the propagation of the electromagnetic field in the vacuum between the gratings, and $\mathcal{R}$ are the gratings' reflection matrices. The arrows under the reflection and propagation matrices indicate the direction of propagation of light and their expression can be obtained using rigorous coupled wave approaches (RCWA) as in classical photonics~\cite{Busch07}. The propagation matrices are diagonal in a plane-wave, Rayleigh basis (see for example~\cite{Davids10} for explicit expressions). All these matrices are evaluated at the Matsubara imaginary frequencies $\omega_l={\rm i} \xi_{l}={\rm i} 2\pi l k_{B}T/\hbar$~\cite{Matsubara55}, and the primed sum indicates that the $l=0$ term has half weight. A particular example of this geometry is represented in Fig.~\ref{gratingPressure}, where the depth $h$ of one of the gratings was reduced to zero to recover a plane. For a grating structure with specific geometrical parameters and comprised by a metal described using the Drude model, the predictions corresponding to \eqnref{pressure} are reported in  Fig.~\ref{gratingPressure}.
At short separations, due to the finite grating conductivity, the pressure scales as $\propto a^{-3}$.  At large separations, the pressure tends towards the value $\zeta(3)k_{B}T/(8\pi a^{3})$, which is the same limiting behavior for the Casimir pressure predicted for the plane-plane configuration using the Drude model.

Despite systems involving `simple' one-dimensional grating structures have been actively investigated both theoretically~\cite{Buscher04,Lambrecht08a,Lussange12,Intravaia12a,Intravaia13,Messina15,Antezza20} and experimentally{~\cite{Chan08,Chiu:2009fqu,Chiu:2010ybt,Bao10,Intravaia13,Wang21}, some disagreements between predictions and measurements of the corresponding Casimir pressure remain. For example, an experiment reported in Ref.~\cite{Intravaia13} measuring the Casimir force between a gold sphere and a one-dimensional gold grating at finite temperature has shown that the Casimir force can be tailored in a nontrivial way by modifying the grating's period~\cite{Intravaia13}. Conversely to comparable measurements involving a dielectric grating~\cite{Chan08,Bao10,Wang21}, however, theoretical predictions and experimental results do not agree, indicating once again the possibility that when metals are involved something in the physics of the system still needs to be understood. 

In the right panel of Fig.~\ref{gratingPressure} we depict the theoretical predictions for the Casimir pressure between a plane and a grating with dimensions very similar to that used in Ref.~\cite{Intravaia13} as well as the designated working range of the \cannex{} setup. The distance range as well as the value of the pressure measured for the largest plane-sphere separation in the experiment reported in Ref.~\cite{Intravaia13} are also represented, showing that \cannex{} has the potential to inspect a complementary regime. Specifically, the device's accuracy of $1$ nPa could allow to investigate the pressure behavior within a range of distances corresponding to the transition to the thermal regime. This is expected to occur for distances of the order of $\lambda_{T}=\hbar c/ (k_{B}T) \sim 7.6$ $\mu$m (green shadowed region in Fig.~\ref{gratingPressure}), way above the largest separation considered in many experiments. Shorter separations could be investigated using the same setup with a slightly more rigid sensor. This would reduce the sensitivity but as the Casimir forces in this distance range scale as $a^{-n}$ with $n$ between 3 and 4, while other disturbing effects, such as patches or electrostatics scale with $2\leq n<4$ the precision of the measurement would not be reduced.

\subsubsection{The thermodynamic state of the system: Configurations out of thermal equilibrium }

The Lifshitz theory of Casimir interactions assumes that the whole system is at thermal equilibrium at temperature $T$. Recent investigations have shown, however, that when nonequilibrium configurations are taken into account, interesting phenomena can occur~\cite{Antezza06,Volokitin07,Reiche22}.
Out-of-equilibrium configurations can be realized with different expedients,
including temperature gradients~\cite{Antezza06,Antezza08,Obrecht07,Messina11a,Klimchitskaya:2019nzu}, moving objects~\cite{Volokitin07,Reiche22} and also scenarios where external lasers act on a system initially in thermal equilibrium~\cite{Bartolo16,Fuchs18a}. In many experiments, nonequilibrium physics is more the rule than the exception.
In particular, the presence of different temperatures in the system can considerably affect the Casimir force's behavior, giving rise to repulsive interactions and different power-law dependencies~\cite{Antezza06,Obrecht07,Messina11a,Klimchitskaya:2019nzu}. In addition to providing alternative ways to tailor Casimir forces, nonequilibrium configurations also offer opportunities to differently investigate the interplay between the Casimir interaction, the material's optical properties, and the system geometry, possibly adding new relevant information for solving some of the issues mentioned above.

\begin{figure}[ht]
\centering
\includegraphics[width=\textwidth]
{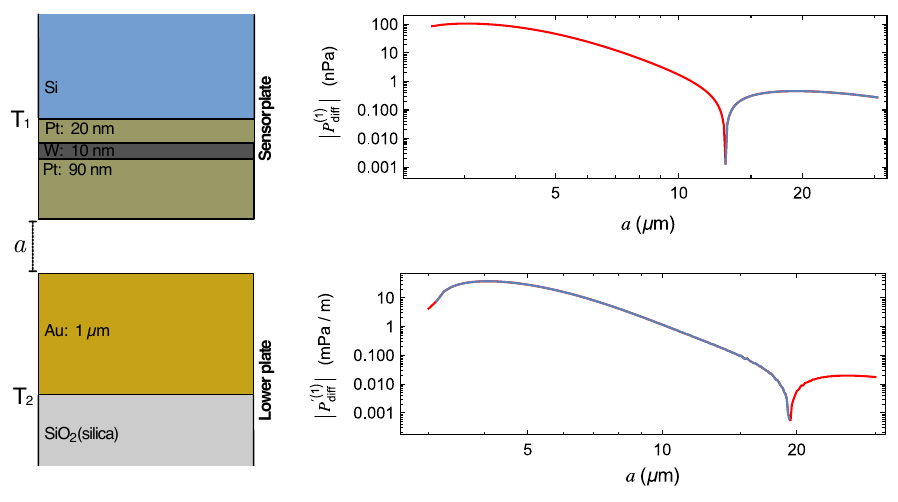}
\caption{(Left) Schematic configuration of the two planar multilayer structures which will be used in the \cannex{} setup for interfacial measurements (not on scale). The lower plate is made by a \SI{1}{\micro\metre} thick gold layer over a silica substrate. The upper/sensor plate is made of a platinum/tungsten/platinum multilayer deposited over a silicon structure. For describing the metals we use the Drude model [see~\eqnref{eq:drude}] with the following parameters. Gold: $\Omega_{\rm Au}=8.39$ eV, $\gamma_{\rm Au}=43.4$ meV~\cite{Intravaia13}. Platinum: $\Omega_{\rm Pt}=5.48$ eV, $\gamma_{\rm Pt}=86.5$ meV~\cite{Mendoza-Herrera17}. Tungsten: $\Omega_{\rm W}=6.41$ eV, $\gamma_{\rm W}=60.4$ meV~\cite{Ordal85}. For simplicity, we described the silicon and silica layers using the same dielectric function described in terms of the Lorentz model [see~\eqnref{eq:lorentz}] with the following parameters: $\epsilon_{0}=11.87$, $\epsilon_{\infty}=1.035$, $\Omega_0=4.346$ eV and $\Gamma=43.5$ meV~\cite{Pirozhenko08}.
(Right) Differential pressure $P^{(1)}_{\rm diff}(a)$ [see \eqnref{DiffPressure}] (top) and its gradient (bottom) corresponding to out-of-equilibrium configurations where $T_{1}=T_{3}=T_{\rm eq}=293$ K while the lower plate's temperature is in one case at temperature $T_{2}=T_{\rm eq}+\Delta T_{2}$ and $T_{2}=T_{\rm eq}-\Delta T_{2}$ in the other case. The value of $\Delta T_{2}$ is taken to be $10$ K corresponding to the temperature difference which can be obtained in \cannex{}. The red curves indicate a negative difference while the blue curves describe positive ones.
\label{fig:OoEPressure}}
\end{figure} 

\begin{figure}[ht]
\centering
\includegraphics[width=0.45\textwidth]
{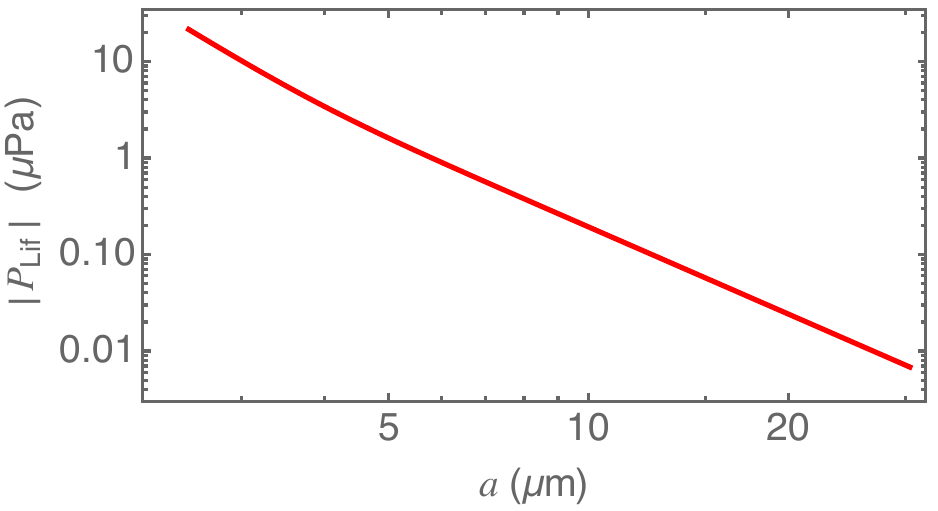}
\hspace{0.4cm}
\includegraphics[width=0.47\textwidth]
{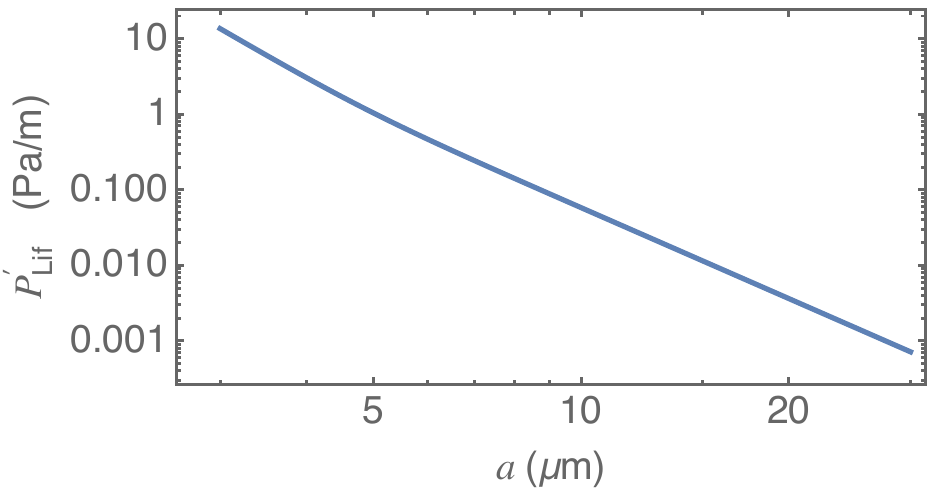}
\caption{Pressure (left) and its gradient (right) in equilibrium at temperature $T=T_{\rm eq}=293$ K for the configuration depicted in Fig.~\ref{fig:OoEPressure}.
The pressure is negative (attraction) while the its gradient is positive.
The values of both quantities are evaluated using the Lifshitz formula reported in \eqnref{Elifshitz} (cfr. Appendix~\ref{thirdterm}) and the material parameters reported in Fig.~\ref{fig:OoEPressure}.
\label{fig:EqPressure}}
\end{figure} 

The high symmetry of the plane-plane configuration has allowed in Ref.~\cite{Antezza08} for a detailed calculation of the thermal nonequilibrium Casimir pressure acting on the inside faces of two planar structure configuration. As in the Lifshitz formula in ~\eqnref{Elifshitz}, the planes can be characterized using the corresponding reflection coefficients $r^{\sigma}_{i}$ ($i=1,2$). If each of the planar structures is assumed to be \textit{locally} in thermal equilibrium at the temperature $T_{i}$ within an environment that is kept at temperature $T_{3}$, the total Casimir pressure on the structure $i$ can be written as follows~\cite{Klimchitskaya:2019nzu} 
\begin{equation}
\label{totalPneq}
P^{(i)}(a,T_{1},T_{2})=\frac{1}{2}\sum_{i=1,2}\left[P_{\rm Lif}(a,T_{i})+\frac{4\sigma_{\rm SB}}{3{\rm c}}T_{i}^{4}\right]
+\Delta P_{\rm neq}(a,T_{1},T_{2})
-\frac{2\sigma_{\rm SB}}{3{\rm c}}(T_{i}^{4}+T_{3}^{4})~,
\end{equation}
where $\sigma_{\rm SB}$ is the Stefan-Bolzman constant. The first term is equivalent to the average equibrium pressure predicted by the Lifshitz formula and the Stefan-Boltzman law evaluated at the two different plates' temperatures. The last term is the pressure of the environmental radiation on the plate $i$ (for both plates we assumed that the external surfaces to be blackened~\cite{Bimonte11b,Klimchitskaya:2019nzu}). 
The second term, $\Delta P_{\rm neq}$, is a pure nonequilibrium contribution: It can be written as the sum of two contributions arising from evanescent and propagating waves, respectively. Moreover, $\Delta P_{\rm neq}$ is nonzero only if the two planar structures are different such that $r^{\sigma}_{1}\neq r^{\sigma}_{2}$ and it is odd if the plates' temperatures are swapped $\Delta P_{\rm neq}(a,T_{1},T_{2})=-\Delta P_{\rm neq}(a,T_{2},T_{1})$ (see Appendix~\ref{thirdterm} for more details).

The direct connection between the detailed expression for $P^{(i)}(a,T_{1},T_{2})$ and their reflection coefficients (from the inside of the resulting cavity) allows for some flexibility in the description of the planar structure and, in particular, for the consideration of multilayered structures~\cite{Zhang05,Yariv83}. An example is given on the left of~\figref{fig:OoEPressure} which represents a typical configuration used in the \cannex{} setup.
A possible measurement scheme involves the upper plate which is kept at equilibrium with the surrounding environment, i.e. $T_{1}=T_{3}=T_{\rm_{eq}}=293$ K, while the lower plate is cooled by $\Delta T_{2}=10$ K during a first measurement campaign and then warmed of the same quantity during a second campaign. According to~\eqnref{totalPneq}, the difference between the two sets of measurements considering the pressures acting on \cannex{}'s sensor plate is given by
\begin{equation}
\label{DiffPressure}
P^{(1)}_{\rm diff}(a)=P^{(1)}(a,T_{\rm eq},T_{\rm eq}+\Delta T_{2})-P^{(1)}(a,T_{\rm eq},T_{\rm eq}-\Delta T_{2}).
\end{equation} 
Since \cannex{} can simultaneously measure both pressure and pressure gradient, on the right side of Fig.~\ref{fig:OoEPressure}, we have plotted the prediction corresponding to these two quantities for the differential measurement described above and in relation to the material configuration on the left side of~\figref{fig:OoEPressure}. For comparison, in Fig.~\ref{fig:EqPressure} we also report the corresponding equilibrium values ($T_i=T_{\rm eq}=293$ k) for both the pressure and its gradient calculated using the Lifshitz formula in \eqnref{Elifshitz}.
Notice that over a range of $3 - \SI{30}{\micro\metre}$ we can predict a maximal value for $P^{(1)}_{\rm diff}(a)$ of about $-0.1$ mPa for a distance $a=3$ to \SI{4}{\micro\metre} and a change in sign from negative ($P^{(1)}(a,T_{\rm eq},T_{\rm eq}+\Delta T)<P^{(1)}(a,T_{\rm eq},T_{\rm eq}-\Delta T_{2})$) to positive ($P^{(1)}(a,T_{\rm eq},T_{\rm eq}+\Delta T_{2})>P^{(1)}(a,T_{\rm eq},T_{\rm eq}-\Delta T_{2})$) around $a=\SI{13}{\micro\metre}$. This means that, for sufficiently short distances, the pressure measured by the sensor plate when the lower plate is warmer than the environment is larger than the corresponding pressure measured for a plate which is cooler than the environment. This balance, however, changes as a function of the separation between the plates. Coherently, the plot on the bottom right corner of~\figref{fig:OoEPressure} shows that pressure gradient, in the range of distance considered in our analysis, changes sign between $3$ and $4$ $\mu$m and again around \SI{20}{\micro\metre}.


\subsection{Scalar Dark Energy}
\label{sec:de}
A prominent approach to solving the cosmological constant problem proposes the existence of new hypothetical scalar fields. However, these scalar fields typically introduce so-called fifth forces. Since such additional forces are tightly constrained by ongoing high-precision experiments, these scalar fields must incorporate some kind of `screening mechanism' to avoid conflict with current experimental results. Several such screening mechanisms have been suggested, such as the chameleon~\cite{Khoury:2003rn,Khoury:2013tda}, K-mouflage~\cite{Brax:2012jr,Brax:2014wla}, Vainsthein~\cite{Vainshtein:1972sx} and Damour-Polyakov~\cite{Damour:1994zq} mechanisms. All these mechanisms have in common that the fifth force is suppressed in high density environments. For this reason, high-precision vacuum experiments such as \cannex{} are ideal tools to probe these hypothetical forces. 

Investigations in this article cover the environment-dependent dilaton~\cite{Brax:2010gi,Sakstein:2014jrq}, symmetron~\cite{Cronenberg:2018qxf,Brax:2017hna,Pitschmann:2020ejb} and chameleon field theories~\cite{Burrage:2017qrf}. Notably, the self-interaction potential of the dilaton finds its theoretical origin in the strong coupling limit of string theory~\cite{Gasperini:2001pc,Damour:2002nv,Damour:2002mi}. The corresponding screening mechanism is highly sensitive to the parameter values and the corresponding behavior has been investigated in detail in~\cite{Fischer:2023koa}. In contrast, symmetrons, resembling the Higgs, employ spontaneous symmetry breaking to realize a screening mechanism. In low density regions, the field is in its spontaneously broken phase and hence acquires a non-vanishing vacuum expectation value (VEV) resulting in a fifth force. However, in high density regions, the symmetry is restored and the fifth force vanishes. Still another screened scalar field theory is the chameleon with a screening mechanism, which increases the mass in dense environments (see e.g.~\cite{Burrage:2017qrf,Brax:2018iyo} for reviews concerning the symmetron and chameleon field). 

While the chameleon and the symmetron field have been constrained by several experiments, such as atomic interferometry~\cite{Burrage:2014oza,Burrage:2015lya}, Eöt-Wash experiments~\cite{Upadhye:2012rc}, gravity resonance spectroscopy~\cite{Brax:2017hna,Cronenberg:2018qxf,Jenke:2020obe}, precision atomic measurments~\cite{Brax:2022olf} and others ~\cite{Burrage:2017qrf,Brax:2018iyo}, more recent investigations on the dilaton model have so far provided only constraints by gravity resonance spectroscopy, Lunar Laser Ranging and neutron interferometry~\cite{Fischer:2023koa,Fischer:2023eww}.
Concerning \cannex{}, prospective constraints have been derived for either of these fields~\cite{Fischer:2023koa,Almasi:2015zpa,Sedmik:2021iaw}. However, these previous analyses suffer from various shortcomings, e.g. the chameleon analysis has neglected the vacuum region above the setup's movable mirror in the calculation of the induced pressure. Furthermore, the chameleon parameter $\Lambda$ has been fixed to the specific value of 2.4 meV. As of now, pressure gradients have not been considered and investigations related to chameleons and symmetrons have not taken variations in vacuum pressure and plate separation into account. Herein, the most rigorous and complete investigation, closing these previous gaps, has been carried out.

\subsubsection{Theoretical background}

The effective potential of the scalar fields considered herein is given by
\begin{align}
    V_{\text{eff}}(\phi; \rho) = V(\phi) + \rho A(\phi),
\end{align}
where $V(\phi)$ is the self-interaction potential and $A(\phi)$ the `Weyl-factor' providing the coupling to the ambient matter density $\rho$. For all investigated models $A(\phi) \simeq 1$ holds. The dilaton (D), symmetron (S), and chameleon (C) models are defined by~\cite{Burrage:2017qrf,Brax:2018iyo}
\begin{align}
\begin{split}
    V_D(\phi) &= V_0\, e^{-\lambda_D \phi /m_{\text{pl}}}\>,\\
    V_S(\phi) &= - \frac{\mu^2}{2}\,\phi^2 + \frac{\lambda_S}{4}\,\phi^4\>,\\
    V_C(\phi) &= \frac{\Lambda^{n+4}}{\phi^n}\>,
\end{split}
\end{align}
together with the Weyl-factors
\begin{align}
\begin{split}
    A_D(\phi) &= 1 + \frac{A_2}{2} \frac{\phi^2}{m^2_{\text{pl}}}\>,\\
    A_S(\phi) &= 1 + \frac{\phi^2}{2M^2}\>, \\
    A_C(\phi) &= e^{\phi / M_c} \simeq 1 + \frac{\phi}{M_c}\>.
\end{split}
\end{align}
The dilaton field is characterized by three parameters, i.e. $V_0$ an energy scale associated with DE, $\lambda_D$ a numerical constant, and $A_2$ a dimensionless coupling parameter. Then, $m_{\text{pl}}$ denotes the reduced Planck mass. Furthermore, the symmetron parameters are given by the tachyonic mass $\mu$, a dimensionless self-coupling constant $\lambda_S$, and $M$ as a coupling constant to matter with a dimension of a mass. Finally, for chameleons $n \in \mathbb{Z}^+ \cup 2\mathbb{Z}^-\setminus$\{-2\} determines the power of the self-interaction potential, $\Lambda$ defines an energy scale that is sometimes related to DE and $M_c = m_{\text{pl}}/\beta$ is a coupling constant with dimension of a mass. To justify the neglect of any higher order couplings, the analysis herein is restricted to 
\begin{align}
    \frac{A_2}{2}\frac{\phi^2}{m^2_{\text{pl}}}\>,\> \frac{\phi^2}{2 M^2}\>,\>\frac{\phi}{M_c} \ll 1\>. \label{cutoff}
\end{align}
The resulting equations of motion are given by
\begin{align}
    \Box \phi + V_{\text{eff},\phi}(\phi; \rho) = 0\>,
\end{align}
while the non-relativistic force acting on a point particle with mass $m$ is~\cite{Pitschmann:2020ejb}
 \begin{align}
    \vec{f}_{\phi} = -m \vec{\nabla}\ln A(\phi)\>.
\end{align}
For the analysis herein, the \cannex{} setup is approximated in 1 dimension along the $z$-axis as follows. The fixed lower mirror is located at $z<0$ with density  $\rho_M = \SI{2514}{kg/\metre^3}$, while the movable upper mirror with density $\rho_M$  and thickness $D = \SI{100}{\micro\metre}$ is located at $a<z<a+D$ with $\SI{3}{\micro\metre}\leq a \leq \SI{30}{\micro\metre}$. Between both mirrors and above the upper mirror vacuum prevails with adjustable density of $\SI{5.3e-12}{kg/\metre^3}\leq \rho_V\leq \SI{0.0026}{kg/\metre^3}$. 
To justify the neglect of the vacuum chamber above the upper plate, an interaction range cut-off at 1 mm has been applied in our analysis. For even greater interaction ranges, the matter content of the vacuum chamber induces a pull on the upper plate thereby effectively lowering the pressure on the upper plate. Hence, the force on the upper mirror is given by~\cite{Brax:2022uyh}
\begin{align}
    \vec{f}_\phi = - \rho_M \int_{-\infty}^{\infty}dx \int_{-\infty}^{\infty}dy \int_{a}^{a+D}dz\,\partial_z\ln A\big(\phi \big)\, \vec{e}_z\>,
\end{align}
and the pressure in $z$-direction on the movable mirror is therefore
\begin{align}
    P&=\rho_M\,\big(\ln A(\phi(a))-\ln A(\phi(a+D)\big) \nonumber\\
&\simeq \rho_M\,\big(A(\phi(a))-A(\phi(a+D)\big)\>. \label{press}
\end{align}
 If the field reaches its potential minimum value $\phi_M$ inside the upper mirror, the latter expression can be simplified further to~\cite{Fischer:2023koa}
\begin{align}
P = \frac{\rho_M}{\rho_M-\rho_V}\,\big(V_{\text{eff}}(\phi_V,\rho_V)-V_{\text{eff}}(\phi_0,\rho_V)\big)\>, \label{newpress}
\end{align}
where $\phi_0:= \phi(a/2)$ is the value of the scalar field in the middle between both plates. This assumption, however, is not very restrictive, since the screening mechanism typically suppresses the field inside the mirror such that the field can effectively reach its potential minimum value. It has been checked explicitly that this assumption is actually satisfied for parameter values where limits were set. In order to obtain $\phi_0$, the following differential equation has to be solved
\begin{align}
    \frac{d^2\phi}{dz^2} - V_{\text{eff},\phi}\big(\phi(z),\rho(z)\big) = 0\>.
\end{align}
Since the field effectively reaches its potential minimum values inside both mirrors, $\phi(z) = \phi_M$ has been set as a boundary condition deep inside the mirrors. For some cases analytical solutions to this equation exist~\cite{Brax:2017hna,Pitschmann:2020ejb,Ivanov:2016rfs}. However, for the new limits obtained herein, this equation has been solved numerically. Whenever possible, a comparison with analytical solutions has been performed as an additional check. This allowed the reliable computation of the pressure as a function of the plate distance as well as the vacuum density. Pressure gradients can straightforwardly be computed by using
\begin{align}
        \partial_aP \simeq \frac{P(a+\delta) - P(a-\delta)}{2\delta}\>,
\end{align}
for small enough $\delta$.

\subsubsection{Dilaton constraints}

\begin{figure}[ht]
\centering
\includegraphics[width=\textwidth]
{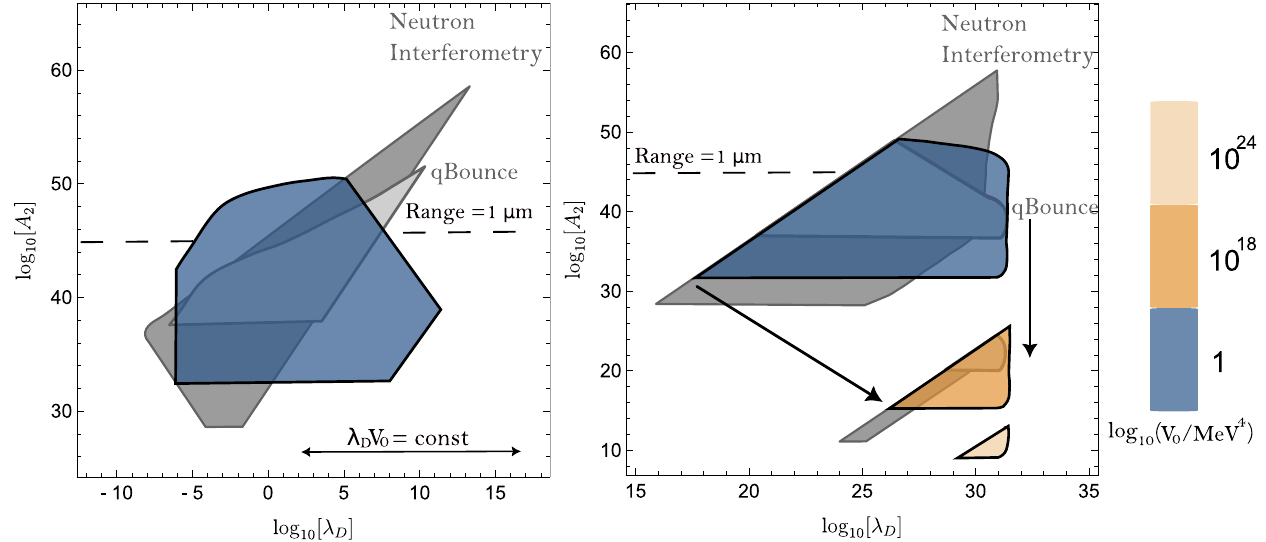}
\caption{Prospective \cannex{} limits on dilaton interactions are shown in color, alongside already existing constraints from $q$Bounce and neutron interferometry (assuming the fermi-screening approximation~\cite{Fischer:2023koa}). The combined constraints from pressure and pressure gradient measurements are plotted. The parameter space of the dilaton field naturally falls into two regimes. \textit{Left:} For small values of the parameter $\lambda_D$ the model has an additional parameter symmetry, such that the physics only depend on the product $\lambda_D V_0$ rather than $\lambda_D$ or $V_0$ individually. Therefore, the shape of the excluded parameter areas remains the same for increasing $V_0$, but only shifts towards lower values of $\lambda_D$ \label{fig:Dilaton}. \textit{Right:} For large values of $\lambda_D$, the dilaton approximately depends only on $A_2 \ln\big(V_0/\rho\big)$, but not on their individual values. Therefore, the excluded parameter areas shift towards lower $A_2$ for increasing $V_0$ without changing their shape. However, in contrast to the small $\lambda_D$ regime, the areas are cut by an ever stronger cut-off. This is indicated by the arrows.}
\end{figure}
The resulting constraints for the dilaton field theory are shown in Fig.~\ref{fig:Dilaton}. \cannex{} will indeed be able to probe parts of the dilaton parameter space that have not been excluded by existing experiments. However, adding pressure gradients to the existing analysis does not improve the constraints that can be obtained with \cannex{}.

\subsubsection{Symmetron constraints}

\begin{figure}[ht]
\centering
\includegraphics[width=0.7\textwidth]
{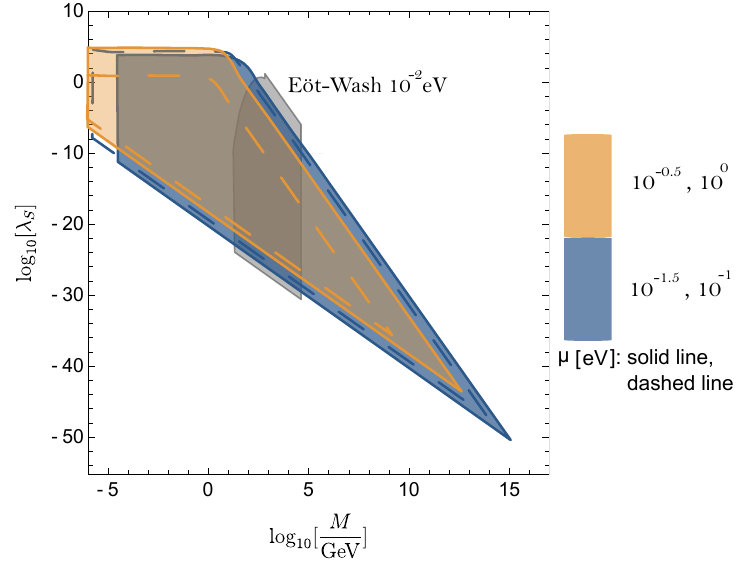}
\caption{Prospective constraints on symmetron interactions from \cannex{} are shown in color. The colored areas refer to constraints for $\mu \in \{10^{-1.5},\ 10^{-0.5}\}$ eV, the colored dashed lines enclose the constraints for $\mu \in \{10^{-1},\ 10^{0}\}$ eV, as indicated in the legend. Only the combined constraints from pressure and pressure gradient measurements are shown, alongside already existing constraints. \label{fig:Symmetron}}
\end{figure} 
The resulting constraints for the symmetron field theory are shown in Fig.~\ref{fig:Symmetron}. 
For too small values of $\mu$ the field vanishes entirely and with it the induced pressure as well. This happens approximately for~\cite{Brax:2017hna}
\begin{align}
\label{eq:SIQ}
    \sqrt{\mu^2-\frac{\rho_V}{M^2}}\,a < \frac{\pi}{2}\>.
\end{align}
For too large $\mu$ values, however, the force between the plates gets very weak. Hence, \cannex{} can only probe a small interval of $\mu$ values. It has been found that in some cases pressure gradients provide better constraints than the pressure itself and that the plate separation has a large impact on the limits. The analysis herein significantly improves on the previous analysis in~\cite{Sedmik:2021iaw}. Specifically, for $\mu = 1$ eV, corresponding roughly to an interaction range of \SI{0.2}{\micro\metre}, the \cannex{} limits have previously been underestimated by a factor of $\sim 10^{20}$ on the $\lambda_S$ axis, since a plate separation of \SI{10}{\micro\metre} was assumed. Clearly, a smaller plate separation of \SI{3}{\micro\metre} yields an enormously stronger pressure and consequently better constraints. Due to the same reason, previous limits for $\mu = 10^{-0.5}$ eV have also been underestimated by several orders of magnitude. Based on \eqnref{eq:SIQ}, in combination with a value of $a= \SI{10}{\micro\metre}$, the conclusion was drawn in~\cite{Sedmik:2021iaw}, that \cannex{} can probe only parameter values $M> 10^2$ GeV for $\mu = 10^{-3/2}$ eV resulting in weak limits. However,  increasing $a$ to \SI{20}{\micro\metre} removes this constraint, and more substantial limits with $M > 10^{-4.5}$ GeV can be obtained, resulting in significant improvements with respect to existing constraints.
Indeed, \cannex{} will be able to improve upon existing table-top experiment constraints by several orders of magnitude. 

\subsubsection{Chameleon constraints}

\begin{figure}[ht]
\centering
\includegraphics[width=0.7\textwidth]
{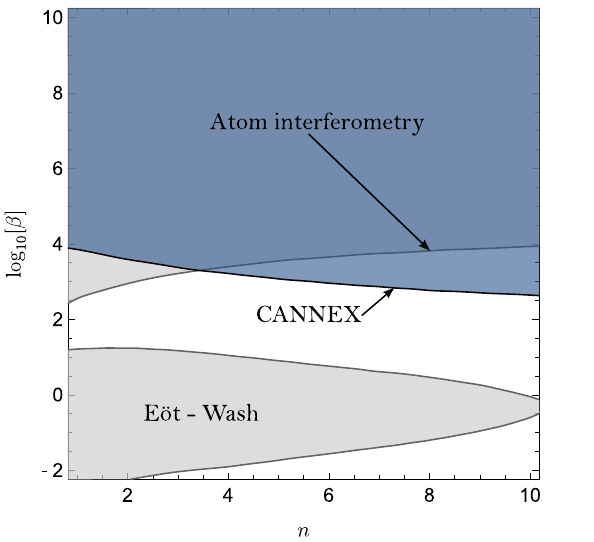}
\caption{The blue area shows the combined prospective constraints of pressure and pressure gradient measurements on chameleon interactions resulting from \cannex{}. The parameter $\Lambda$ has been fixed to the DE scale of 2.4 meV.\label{fig:Chameleon}}
\end{figure} 
Since each value of $n$ is typically considered as a separate chameleon model, the analysis herein has been restricted to two cases. The most commonly studied model is $n=1$ and hence limits have been computed for $n=1$ and varying $\Lambda$. However, within the current limits of the applied theoretical analysis, no new parts of the parameter space can be probed using \cannex{}. Nevertheless, fixing $\Lambda = 2.4$ meV to the DE scale and varying $1 \leq n\leq 10$, which is also commonly studied, will indeed result in narrowing the gap between existing limits as shown in Fig.~\ref{fig:Chameleon}. 

\section{Discussion}
\label{sec:discussion}
\cannex{} has completed its design phase and is about to be realized with first results expected in 2024. It is the first experiment to perform highly accurate measurements of both interfacial and gravity-like forces and force gradients in the distance regime 3--\SI{30}{\micro\metre} with truly plane parallel plates. This geometry increases the sensitivity to distance-dependent forces by several orders of magnitude with respect to curved interacting surfaces used in most other experiments. High accuracy naturally demands for control of various disturbing effects. We have designed and (partially) tested thermal control at the (sub-)mK precision level both in thermal equilibrium and with the two interacting plates being out of thermal equilibrium by \SI{10}{\celsius}. We also designed a six-axis passive seismic attenuation system, in-situ surface charge and impurity removal by UV irradiation and Ar ions, purely optical detection systems, and an in-situ Kelvin probe / AFM setup to characterize the surfaces. Importantly, our calibration procedures rely on references that can be traced to metrological standards (wavelengths, voltages, frequencies).

In the present article, we give a final update on the design and measurement procedures, on the basis of which we compute a detailed update of the detection error and seismic disturbances. Using specifications of and noise measurements measurements with the actual devices, we find that the error at large separation $a\gtrsim\SI{15}{\micro\metre}$ can realistically be reduced by factors $2$ and $30$ in the pressure and its gradient, respectively, with respect to the previous error budget in Ref.~\cite{Sedmik:2021iaw}.

\cannex{} can be operated in different configurations. In interfacial configuration, the two plates directly face each other, which allows us to measure Casimir forces, and hypothetical screened scalar DE forces. Measuring the former ones at the percent level at separations both smaller and larger than the thermal wavelength will allow us for the first time to probe the transition from predominantly quantum mechanical origin to thermal origin in the Casimir force at high accuracy. The respective data may lead to further insight regarding longstanding problems regarding the role of dissipation and locality in the description of the dielectric response of metals. \cannex{} could also perform the first quantitative measurements of Casimir forces out of thermal equilibrium, thereby testing the respective theory. Eventually, the plates can easily be modified by structuring. Together with the control of parallelism at the \si{\micro\radian} level, gratings or cylinders at arbitrary angles could be investigated, thereby generating high-precision data that can be used to further verify theoretical approaches currently disagreeing with measurement results for such geometries. While the Casimir force is a worthwhile subject to study, it also poses a nuisance if we aim to measure screened scalar DE forces, which have a similar distance dependence but in comparison several orders of magnitude lower strength. An electrostatic shield would entirely block these interactions. For this reason, we adapt our measurement procedure by remaining in interfacial configuration at the same surface separation but changing the ambient pressure of Xe gas. In the presence of the gas, electrostatic, Casimir and gravitational forces will only show a negligible and calculable increase, while screened DE forces would decrease in strength. In a relative measurement with and without Xe, our sensitivity to such hypothetical forces is maximized. Here we present updated prospective exclusion graphs for the most prominent representatives of screened scalar fields: dilatons, symmetrons, and chameleons. The new calculations properly take into account the finiteness and geometry of the interacting objects, the formation of the field within the vacuum chamber, and the validity limits of the theory -- aspects that have mostly been neglected in the literature. For all three scalar fields, even considering only the most conservative error budget, \cannex{} will be able to extend present limits by several orders of magnitude.

The second possible configuration is the one of Cavendish, where a thin flat conducting shield is added between the two plates in order to remove electrostatic and Casimir forces from the balance. In this configuration, volume-sourced forces, such as gravity or hypothetical fifth forces between fermions in the two plates can be measured with high sensitivity. As we re-confirm (and slightly improve) the previous error budget and prospective limits on a variety of such forces have been presented recently, we do not update these data here but refer to the literature~\cite{Sedmik:2021iaw}. With this in mind, \cannex{} will be able to measure gravity (and thereby Newton's constant $G$) at the 10\% level in a distance regime down to \SI{10}{\micro\metre} with active masses of roughly \SI{30}{\milli\gram}, thereby probably exceeding recent experiments with torsion balances and spherical objects probing in this direction. We remark that, since the thick metal coatings on the plates of \cannex{} have densities that exceed those of the carrier material by an order of magnitude, the effective separation between our masses sourcing the gravitational interaction lies close to the actual surface separation -- in stark difference to spherical objects.
\vspace{6pt}
\section*{Author contributions}
Conceptualization, R.I.P.S., H.F., M.P. and F. I.; methodology, R.I.P.S. and M.P.; software, H.F., R.I.P.S. and R.S.; design and simulation: I.G., A.B., R.S., H.H. and R.I.P.S., formal analysis, H.F., H.H. and R.I.P.S.; writing ---original draft preparation, H.F., R.S., H.H., F.I., and R.I.P.S.; writing---review and editing, R.I.P.S., M.P., R.S., I.G. and A.B. visualization, R.S. and R.I.P.S.; supervision, M.P. and R.I.P.S.; project administration, R.I.P.S.; funding acquisition, R.I.P.S. and M.P. All authors have read and agreed to the published version of the manuscript.
\section*{Acknowledgements}
This research was funded by the Austrian Science Fund (FWF) under grant No. P 36577-N and P 34240-N. The authors thank R. Gergen, H. Matusch, A. Pelczar, T. Dokulil, and P. Suprunowicz for technical support, M. Maichanitsch for analyses of the SAS design, and H. Abele for administrative support. The previous version of this experiment was funded by the Netherlands Organization for Scientific Research (NWO) and donations via the Cannex crowd funding campaign by ASML and others.
\section*{Abbreviations}
The following abbreviations are used in this manuscript:\\

\noindent 
\begin{tabular}{@{}ll}
AC & Alternate Current\\
AFM & Atomic Force Microscope\\
AM & Amplitude Modulation\\
ATI & ATomInstitut (of TU Wien)\\
CANNEX & Casimir And Non-Newtonian force EXperiment\\
COBS & Conrad OBServatory\\
CKM & Cabibbo-Kobayashi-Maskawa matrix\\
CP & Charge Parity\\
DC & Direct Current\\
DE & Dark Energy\\
DM & Dark Matter\\
ES & Electrostatic\\
FEM & Finite Element Method\\
FM & Frequency Modulation\\
GAS & Geometric Anti-Spring\\
GR & General Relativity\\
KPFM & Kelvin Probe Force Microscopy\\
$\Lambda$-CDM & Cold Dark Matter model with cosmological constant $\Lambda$\\
PE & Peltier Element\\
QCD & Quantum ChromoDynamics\\
QED & Quantum ElectroDynamics\\
LED & Light-Emitting Diode\\
LVDT & Linear Variable Differential Transformer\\
MC & Monte Carlo (simulation)\\
NHNM & New High Noise Model\\
NLNM & New Low Noise Model\\
PLL & Phase-Locked Loop\\
SAS & Seismic Attenuation System\\
SM & Standard Model (of particle physics\\
SNR & Signal to Noise Ratio\\
UHV & Ultra-High Vacuum\\
UV & UltraViolet\\
WIMP & Weakly Interacting Massive Particle\\
\end{tabular}

\appendix
\section[\appendixname~\thesection]{Details of the error budget}
\label{app:error_details}
In the following, we give the models used for the calculation of various specific detection errors. For easier reference, we give them in a listed format below in their original units, while their effects on the respective measurement are listed in \tabref{tab:dc-error}, \ref{tab:ac-error},\ref{tab:df-error}, and \ref{tab:f-error} in the main text.\\
As a general model for drift estimation, we assume a generic diurnal sinus model with $A_T=\SI{0.1}{\celsius}$ amplitude,
\begin{align}
TD(t)= A_T\sin \frac{2\pi t}{24\times3600\,{\rm s}}\,,
\label{eq:drift_model}
\end{align}
which over-estimates the actual temperature variation at COBS and statistically exceeds the error for a normal distribution by up to two orders of magnitude, but serves as a worst-case scenario. \tabref{tab:global_param} lists further parameters assumed throughout the analysis. For RMS quantities $x_\text{RMS}$, we do not use a sharp cutoff at the bandwidth $1/\tau$ but integrate the spectrum $x(\omega)$ up to the maximum frequency $\omega_\text{max}$ in the spectrum with a first-order low-pass $T_I(\omega,\tau)=[1+\ri\omega\tau]^{-1}$, yielding a more realistic estimate
\begin{align}
x_\text{RMS}(\tau)=\left[\int_0^{\omega_\text{max}}\!\frac{{\rm d}\omega}{2\pi}x^2(\omega)|T_I|^2(\omega) \right]^{\frac{1}{2}}\,.\label{eq:rms_integration}
\end{align}
Global parameters of the sensor and measurements are listed in \tabref{tab:global_param}. Note again that all statistical errors where we do not have spectra available are time-averaged with $1/\sqrt{t/\tau}$, where $\tau$ is the time constant of the actual measurement as listed in the table. I. Systematic errors are averaged with $1/\sqrt{N_\text{cal}}$, where $N_\text{cal}$ is the number of calibrations. We furthermore use the following indicators on statistics: `SpI' means that spectral integration is performed according to \eqnref{eq:rms_integration} instead of regular time averaging; `TD' indicates that temperature drift is considered on this error. 
\begin{table}[!ht]
\caption{Global parameters assumed for the error analysis.\label{tab:global_param}}
\centering
\begin{tabular}{l r@{.}l c p{9 cm}}
\toprule
parameter & \multicolumn{2}{c}{value} & unit & description\\
\midrule
$\tau_\text{DC}$ & 2&0 & s &  integration time for a single DC voltage measurement\\
$\tau_\text{AC}$ & 83&0 & s &  lock-in integration time for a single AC amplitude, frequency or phase measurement\\
$m$ & 26&13 & mg & effective dynamic sensor mass\\
$\omega_0$ & 2$\pi\times 9$&8 &s$^{-1}$ & free sensor resonance frequency\\
$ d$ & 500 & & \si{\micro\metre} & nominal sensor cavity size\\
$ A$ & 1&035& cm$^2$ & sensor interaction area\\
\bottomrule
\end{tabular}
\end{table}
\subsection{DC Signals}
\label{app:errors_DC}
\subsubsection*{Statistical errors}
\begin{description}
\item [DAQ noise] $\delta_\text{DAQ} = (\inv{2}+1)\times 10^{-7}\,\text{V}
$ containing the aliasing error from 34470A datasheet (first term). Keysight specifies~\footnote{Private communication.} that the error given in the datasheet is for a temperature range of $\pm1^\circ$C and can be adapted if the real temperature variation is below that. We add \SI{1}{\micro\volt} (second term) to account for noise picked up by cabling, estimated from actual measurements with the device.
\item [Cavity size fluctuations] $\delta d$. We consider the RMS value according to \eqnref{eq:rms_integration} of the measured vertical vibration spectrum [see \figref{fig:seismic_background}] at COBS, the passive SAS $T_{x0x2}(\omega)$, and the sensor response $T_{z_0z}(\omega)$ up to $1/{\tau}\,$Hz with $\tau\geq\tau_\text{DC}$. For $\tau=\tau_\text{DC}$, we have $\delta d=\SI{8.5}{\pico\metre}$.
\item [Detector noise] $\delta V_\text{Det}$. At $\lambda=\SI{1550}{\nano\metre}$, the detectors have a noise level of \SI{0.19}{\pico\watt/\sqrt{\hertz}}, at a total incident flux of \SI{1}{\milli\watt} from the fiber interferometer into the detector (based on laser power and the optical properties of the cavity and fiber). We consider a \SI{1}{\kilo\hertz} bandwidth for the low-pass filter, resulting in \SI{60}{\nano\volt} RMS noise.
\item [Laser power fluctuations] $\delta P_L$. We received actual TLX1 intensity noise spectra from the manufacturer ranging from $1/(3\times3600)\,{\rm s}$ to \SI{10}{\kilo\hertz}. From these data, we determined a temperature correlation coefficient of $4.38\pm0.03\times 10^{-6}\,{\rm K}^{-1}$ but not all of the drift is temperature-related. We thus use the measured Allan deviation as error here. For integration times 2, 83, 1000\,s, using \eqnref{eq:rms_integration}, we obtain RMS relative intensity errors $5.53\times 10^{-4}$, $1.51\times 10^{-5}$, $1.54\times 10^{-6}$, respectively. To the first order, this error is canceled exactly by the normalization in \eqnref{eq:det_error}. Indicator: SpI
\item [Laser bandwidth] $\delta \lambda_\text{BW}$. Given by the datasheet to be \SI{10}{\kilo\hertz} (0.08\,fm) nominally, as the low-frequency limit of the frequency noise.
\item [Laser frequency noise] $\delta \lambda_f$. Derived from manufacturer data of the spectral frequency noise $\delta f$ in the range 3\,Hz--100\,MHz. At lower frequencies, the noise is assumed to stay constant at \SI{6.8}{\mega\hertz/\sqrt{\hertz}}, which is three orders of magnitude larger than the specified linewidth but serves as a worst-case estimate. We convert these data to wavelength noise by $\delta \lambda_f=(\lambda^2/c) \delta f$ for the mean wavelength $\lambda=1590\,$nm after integration over the spectrum as described above, resulting in RMS values \SI{6.28}{\femto\metre}, \SI{1.04}{\femto\metre}, \SI{0.31}{\femto\metre} for $\tau=2\,,\ 83\,,\ 1000\,$s, respectively. Indicator: SpI
\item [Reference cavity signal] $\delta V_\text{R}$. Respective values are obtained from the total $\delta V_\text{DC}$ without seismic vibrations and thermal distance fluctuations, as the reference cavity is a monolithic block made of a material with thermal expansion coefficient $<2\times 10^{-8}\,{\rm K}^{-1}$. We obtain a total $\delta V_\text{ref}=\SI{4.82e-6}{\volt}\,|\,\SI{1.06e-6}{\volt}$ for 1000\,s\,|\,72\,h integration time, indicating the errors for $V_\text{R}(t)$, and $V_\text{R}(0)$,respectively. This error could be reduced in practice, as power fluctuations being the main error here also have a significant temperature dependence.
\end{description}
\subsubsection*{Systematic errors}
\begin{description}
\item [DAQ error] $\sigma V_\text{DAQ}$. We use the temperature drift according to manufacturer specifications $\sigma_\text{DAQ}=(S_A\times \SI{1}{\micro\volt}+\SI{1}{\micro\volt})\times TD(\tau)$ with $S_A=0.55\,$V\,. For longer measurements, we consider a reset of this error by the Keysight 34470A's auto-calibration routing after $\tau=1000\,$s. Indicator: TD
\item [Cavity drift] $\sigma d(t)$ The effective temperature coefficient of $d$ can only be measured, as uncertainties in the material properties lead to rather different values. Considering the actual geometry and materials, we obtain an estimate of $5\times10^{-8}\,$m/K which, together with a preliminary stability 0.1\,mK of the core temperature and $TD(\tau)$ results in the second-strongest error at large $\tau$. Knowing the actual temperature, this error could be removed from the results but we do not consider this possibility in the error budget here. We rather assume that $\sigma d(t)$ can be reset using a $\lambda$-sweep calibration preceding each measurement point, leading to respective statistical averaging and consider $TD(\tau)$ with amplitude $5\,$pm. We add to $\sigma d(t)$ the uncertainty of determination obtained from simulated calibration data. For this purpose, we computed 100 $\lambda$-sweep datasets considering independently randomized $\delta \lambda$, $\delta V_\text{sig}$, $\delta d$, and fixed $\sigma \lambda$ with their respective known statistical widths. The single sweep data are fit to \eqnref{eq:opt_signal} with free parameters $S_A$, $S_B$, $d$ and $\sigma \lambda$. $\sigma d$ is then the standard deviation of all MC results and the mean parameter error (added as systematic errors) of the fits. The same procedure is used for the reference cavity size determination error $\sigma d_\text{ref}$, where we set $\delta d=0$ for data generation. For the computation of the 72\,h reference signal, we assume periodic re-calibration and reset of $\sigma d(t)$ every $500\tau_\text{DC}+\tau_\text{cal}$, with calibration time $\tau_\text{cal}=2800\,$s. Indicator: TD
\item [Wavelength drift] $\sigma \lambda(t)$ is derived from the $1.5\,$GHz accuracy of the TLX1 for a range 10--\SI{40}{\celsius}. As the absolute wavelength can be re-calibrated using the frequency-locked reference laser, we assume for operation at COBS a pessimistic max. error of $\SI{12.6}{\pico\metre}/100$ as amplitude for $TD(\tau)$. This error averages with the number of measurement points. or the computation of the 72\,h reference signal, we assume periodic re-calibration and reset of $\sigma \lambda(t)$ every $500\tau_\text{DC}+\tau_\text{cal}$. Indicator: TD
\item [Reference cavity signal]: Systematic component of $\sigma V_\text{R}=4.7\,|\,\SI{0.79}{\micro\volt}$ for 1000\,s\,|\,72\,h integration time, respectively. Obtained in the same way as $\sigma V_\text{DAQ}$. Indicator: TD
\end{description}
\subsubsection*{Constant errors}
\begin{description}
\item [DAQ error] $\sigma_\text{DAQ} = \SI{0.1}{\micro\volt}$ for the Keysight 34470A offset error, exceeding the specifications from the datasheet.
\item [Reference cavity signal] Constant component of $\sigma V_\text{ref}=\SI{0.1}{\micro\volt}$, similar as for $\sigma_\text{DAQ}$.
\end{description}
Note that constant errors in $d$ and $\lambda$ do not appear as voltage errors due to measurement at quadrature. They are considered in \appref{app:errors:pressure_grad} and \ref{app:errors:pressure}.
\subsection{AC Signals}
\label{app:errors_AC}
\subsubsection*{Statistical errors}
\begin{description}
\item[PLL frequency noise] $\delta f_\text{LI}$. The short-time stability of the lock-in amplifier's phase-tracking based on phase stability $\delta \phi_\text{LI}$ was measured as the RMS value of the phase using a first-order passive RC-lowpass as a device under test over 3\,h, without feedback. This error combines internal electrical noise, aliasing errors, and internal oscillator stability (without an external Rubidium reference clock). We obtained $\delta f_\text{LI}\leq \delta \phi_\text{LI} \omega_0/(4\pi Q)=\SI{1.80}{\nano\hertz}$ (for $Q=10^4$ and $\delta \phi_\text{LI}=\SI{2.4e-4}{\degree}$).
\item[Frequency measurement] $\delta f_\text{PID}$. This noise quantifies the stability of the frequency tracking algorithm of the PLL together with PID feedback. We measured it using the same first-order passive RC lowpass, resulting in $\delta f_\text{PID}=\SI{2.2}{\micro\hertz}$
\item[Signal noise] $\delta f_V$. Voltage noise (containing all error sources described in \appref{app:errors_DC}) can be converted into time jitter of a sinusoidal signal at frequency $\omega$ as explained in the main text in \secref{page:frequency_noise_from_voltage}, resulting in $\delta V_\text{Sig}=7.39\times 10^{-7}\,|\,5.59\times10^{-7}\,$V, $\sigma f_V=3.90\,|\,\SI{0.51}{\nano\hertz}$ for $\tau_{AC}\,|\,1000\,$s integration time and $a=\SI{3}{\micro\metre}$. Indicator: SpI (indirectly, see \appref{app:errors_DC})
\end{description}
\subsubsection*{Systematic errors}
\begin{description}
\item[PLL phase stability] $\sigma f_\text{LI}(t)$. This error quantifies the $0.05\,$ppm/\si{\celsius} drift of the internal oscillator of the lock-in amplifier with temperature, and the respective deviation at COBS. For multiple measurements, we consider periodic re-calibration to average this error. $\sigma f_V(t)=0.88\,|\,\SI{10.5}{\nano\hertz}$ for $\tau_\text{AC}\,|\,1000\,$s integration time, respectively. Indicator: TD
\item[Resonance freq. cal. error] $\sigma \omega_0$. The resonance frequency is calibrated prior to each distance sweep or once per day. We use the combined standard deviation and parameter error obtained from MC simulations of calibration data as described in \secref{sec:setup:omsystem:calibration}. $\sigma \omega_0=1.44\times 10^{-9}\,{\rm s}^{-1}$.
\item[Signal drift] $\sigma f_V(t)$. Drifts of the voltage signal, converted to frequency error as described in \secref{sec:error:detection}. We obtain $\sigma V_\text{Sig}=3.64\times 10^{-7}\,|\,6.69\times 10^{-7}\,$V, $\sigma f_V(t)=1.92\,|\,\SI{3.63}{\nano\hertz}$ for $\tau_\text{AC}\,|\,1000\,$s integration time, respectively and $a=\SI{3}{\micro\metre}$. Indicator: TD (indirectly, see \appref{app:errors_DC})
\end{description}
\subsubsection*{Constant errors}
\begin{description}
\item[PLL phase error] $\sigma f_\text{LI}$. This error reflects the absolute $0.05\,$ppm frequency accuracy of the reference Rubidium atomic clock, applied to the sensor resonance frequency ($f_0=10\,$Hz). 
\item[Resonance freq. cal. error] $\sigma \omega_0$. This error comes from the mean constant offset error seen in our MC simulations. It is caused by non-linearities in combination with other errors, leading to a constant estimation error $\sigma \omega_0=\SI{5.53e-7}{\second^{-10}}$.
\item[Signal error] Constant component of the signal error, amounting to $\sigma f_V=\SI{0.1}{\micro\volt}$ or $0.53\,$nHz (see \secref{sec:setup:omsystem:calibration}).
\end{description}
\subsection{Pressure gradient}
\label{app:errors:pressure_grad}
\subsubsection*{Statistical errors}
\begin{description}
\item[Frequency measurement] $\delta f$. This error is propagated from the AC error described in \appref{app:errors_AC} and amounts for $\tau_\text{AC}$ to $\delta f=1.62\,|\,\SI{0.47}{\micro\volt}$ for $\tau=\tau_\text{AC}\,|\,1000\,$s, respectively at $a=\SI{3}{\micro\metre}$. Indicator: SpI (indirectly, see \appref{app:errors_AC})
\end{description}
\subsubsection*{Systematic errors}
\begin{description}
\item[Frequency measurement] $\sigma f(t)$. This error is propagated from the AC error described in \appref{app:errors_AC}. We obtain $\sigma f=0.71\,|\,\SI{7.76}{\nano\hertz}$ for $\tau=\tau_\text{AC}\,|\,1000\,$s, respectively. Indicator: TD (indirectly, see \appref{app:errors_DC})
\item[Resonance freq. cal. error] $\sigma \omega_0$. This error described already in \secref{app:errors_AC} is considered separately here, as it appears in the expression for the total gradient $\partial_a F$, expressed from \eqnref{eq:frequency_shift}. $\sigma \omega_0=1.44\times 10^{-9}\,{\rm s}^{-1}$.  
\item[mass calibration error] $\sigma m$. We use again the standard deviation and parameter error determined from MC simulations of calibration data (see \secref{sec:setup:omsystem:calibration}). $\sigma m=5.86\times10^{-11}\,$kg.
\end{description}
\subsubsection*{Constant errors}
\begin{description}
\item[Frequency measurement] $\sigma f=\SI{5.13}{\nano\hertz}$ is the constant part of the error propagated from the AC frequency detection.
\item[Resonance frequency error] $\sigma \omega_0=\SI{5.53e-7}{\second^{-10}}$. Mean parameter offset from fits to MC simulation data (see \secref{sec:setup:omsystem:calibration}). 
\item[Mass calibration error] $\sigma m=1.28\times 10^{-11}\,$kg. Mean parameter offset from fits to MC simulation data (see \secref{sec:setup:omsystem:calibration}).
\end{description}
\subsection{Pressure}
\label{app:errors:pressure}
\subsubsection*{Statistical errors}
\begin{description}
\item[Signal fluctuation] $\delta V_\text{Sig}$. Propagated statistical error from the DC signal. Amounts to $\delta V_\text{Sig}=0.74\,|\,\SI{0.10}{\micro\volt}$ for $\tau_\text{AC}\,|\,1000\,$s integration time, respectively at $a=\SI{3}{\micro\metre}$.
\item[Reference signal] $\delta V_\text{0}$. Statistical error of the zero-force reference signal taken at $a_\text{cal}$ (do not confuse with $\delta V_\text{R}$ from the reference cavity). As DC detection is independent of $a$, we use the same models as for $\delta V_\text{Sig}$ described in \appref{app:errors_DC}. $\delta V_0=\SI{0.01}{\micro\volt}$ for $\tau=1000\,$s integration time.
\item[Force gradient] $\delta\partial_a F$. Correcting the spring constant $k$ introduces a dependence on the force gradient. We propagate the corresponding error described in \appref{app:errors:pressure_grad}, resulting in $\delta \partial_a F=32.5\,|\,9.36\,$nN/m for $\tau_\text{AC}\,|\,1000\,$s integration time, respectively.
\end{description}
\subsubsection*{Systematic errors}
\begin{description}
\item[Mass calibration error] $\sigma m=5.86\times10^{-11}\,$kg was described in \appref{app:errors:pressure_grad}.
\item[Resonance freq. error] $\sigma \omega_0=1.44\times 10^{-9}\,{\rm s}^{-1}$. This is the same error described in \appref{app:errors_AC}.
\item[Wavelength error] $\sigma \lambda(t)$. While $\sigma \lambda$ can be measured and brought close to zero by the beat method (see \secref{sec:setup:omsystem:calibration}), it can also be obtained from a fit to a $\lambda$-sweep (see $\sigma d$ above). We use the average parameter uncertainty of the fits combined with the standard deviation of the results using 300 sets of calibration data, resulting in $\sigma\lambda=0.16\,$pm. In addition, we use the known temperature dependence as described in \appref{app:errors_DC}: $\sigma\lambda=\SI{12.6}{\pico\metre}/100\times TD(\tau)$, and add the two uncertainties. Indicator: TD (partially) 
\item [Cavity size determination error] $\sigma d(t)$. Same as described in \appref{app:errors_DC}.
\item[Signal error] $\sigma V_\text{Sig}$. Systematic component of the signal error from \appref{app:errors_DC}. We use $\sigma V_\text{Sig}=0.36\,|\,\SI{0.69}{\micro\volt}$ for $\tau_\text{AC}\,|\,1000\,$s, respectively. Indicator: TD (indirectly, see \appref{app:errors_DC})
\item[Reference signal error] $\sigma V_0$. Systematic error of the zero-force reference, from \appref{app:errors_DC} for $\tau=1000\,$s. $\sigma V_0=\SI{0.69}{\micro\volt}$. Indicator: TD (indirectly, see \appref{app:errors_DC})
\item[Force gradient error] $\sigma\partial_a F$. Systematic error of the synchronous force gradient measurement, considering all errors from \appref{app:errors:pressure_grad}, $\sigma \partial_a F=3.55\,|\,3.55$ for $\tau=\tau_\text{AC}\,|\,1000\,$s, respectively. Indicator: TD (indirectly, see \appref{app:errors:pressure_grad})
\end{description}
\subsubsection*{Constant errors}
\begin{description}
\item[Resonance frequency error] $\sigma \omega_0=\SI{5.53e-7}{\second^{-10}}$. Mean offset from MC simulations, see \appref{app:errors:pressure_grad}. 
\item[Mass calibration error] $\sigma m=1.28\times 10^{-11}\,$kg. Mean offset from MC simulations. 
\item [Wavelength offset error] $\sigma \lambda$. Absolute error of the LLD1530 reference laser from manufacturer data, adjusted for better thermal stability at COBS, as described in \secref{sec:error:detection}. During the experiment, this may turn out to be a systematic error. Conservatively, we consider it to be constant here. $\sigma \lambda=3.4\,$fm.
\item[Signal error] $\sigma V_\text{Sig}$. Propagated constant error of the DC signal $V_\text{Sig}$. $\sigma V_\text{Sig}=\SI{0.19}{\micro\volt}$.
\item[Reference signal error] $\sigma V_0$. Constant error of the zero-force reference signal. $\sigma V_0=\SI{0.19}{\micro\volt}$.
\item[Force gradient error] $\sigma\partial_a F$. Constant part as described in \appref{app:errors:pressure_grad}, amounting to $\sigma\partial_a F=1.16\,$nN/m.
\end{description}
\subsection{Other errors}

The radius of the plates is specified with uncertainty \SI{5}{\micro\metre}. It can be measured with slightly better accuracy. To convert our errors on the force and its gradient to a pressure and pressure gradient, we consider a constant error $A\to A(1+\sigma A)$ with $\sigma A=2.5\times 10^{-3}$ considering the max. deviation on both plates and alignment errors.
\section[\appendixname~\thesection]{Evaluation of the out of thermal equilibrium Casimir pressure.}
\label{thirdterm}

In the expression for the nonequilibrium Casimir pressure given in \eqnref{totalPneq} the pure nonequilibrium term, $\Delta P_{\rm neq}$, can be written as the sum of a contribution arising from the evanescent waves ($\Delta P^{\rm EW}_{\rm neq}$) and a contribution due to propagating waves ($\Delta P^{\rm PW}_{\rm neq}$) \cite{Antezza08,Bimonte11b,Klimchitskaya:2019nzu}.
Considering local and isotropic materials, the two contributions can be conveniently written as
\begin{subequations}
\label{Pneq}
\begin{align}
\Delta P^{\rm EW}_{\rm neq}(a,T_{1},T_{2})
&=-\hbar\int_{0}^{\infty}\frac{d\omega}{\pi}\Delta n(\omega,T_{1},T_{2})\int_{\frac{\omega}{c}}^{\infty}\frac{dk}{2\pi}k \sum_{\sigma} \kappa \frac{\mathrm{Im}[r^{\sigma}_{1}]\mathrm{Re}[r^{\sigma}_{2}]-\mathrm{Re}[r^{\sigma}_{1}]\mathrm{Im}[r^{\sigma}_{2}]}{\vert 1-r^{\sigma}_{1}r^{\sigma}_{2}e^{-2\kappa a}\vert^{2} }e^{-2\kappa a}
\nonumber\\
&=-\frac{\hbar}{2}\int_{0}^{\infty}\frac{d\omega}{\pi}\Delta n(\omega,T_{1},T_{2})\int_{0}^{\infty}\frac{d\kappa}{2\pi}\sum_{\sigma} \kappa^{2} \frac{2\, \mathrm{Im}[r^{\sigma}_{1}(r^{\sigma}_{2})^{*}]e^{-2\kappa a}}{\vert 1-r^{\sigma}_{1}r^{\sigma}_{2}e^{-2\kappa a}\vert^{2} }~,
\end{align}
\begin{align}
\Delta P^{\rm PW}_{\rm neq}(a,T_{1},T_{2})
&=-\frac{\hbar}{2}\int_{0}^{\infty}\frac{d\omega}{\pi}\Delta n(\omega,T_{1},T_{2})\int_{0}^{\frac{\omega}{\rm c}}\frac{dk}{2\pi}k\sum_{\sigma} k_{z} \frac{\vert r^{\sigma}_{1}\vert^{2}-\vert r^{\sigma}_{2}\vert^{2}}{\vert 1-r^{\sigma}_{1}r^{\sigma}_{2}e^{2{\rm i} k_{z} a}\vert^{2} }
\nonumber\\
&=-\frac{\hbar}{2}\int_{0}^{\infty}\frac{d\omega}{\pi}\Delta n(\omega,T_{1},T_{2})\int_{0}^{\frac{\omega}{\rm c}}\frac{dk_{z}}{2\pi}\sum_{\sigma} k_{z}^{2} \frac{\vert r^{\sigma}_{1}\vert^{2}-\vert r^{\sigma}_{2}\vert^{2}}{\vert 1-r^{\sigma}_{1}r^{\sigma}_{2}e^{2{\rm i} k_{z} a}\vert^{2} }~,
\end{align}
\end{subequations}
where it was also assumed that each of the plates is locally in thermal equilibrium at the corresponding temperature.
In the previous expressions, we have used the same conventions and definitions described after \eqnref{Elifshitz}, while  `$^{*}$' indicates the complex conjugate of the corresponding quantity. In the second line of the first equation, given that $\kappa$ is nonnegative over the whole integration range, we changed the variable from $k$ to $\kappa$. Similarly, in the second line of the second equation we performed the variable change $k\to k_{z}=\sqrt{\omega^{2}/c^{2}-k^{2}}={\rm i} \kappa$ ($\mathrm{Im}[k_{z}]\ge 0; \mathrm{Re}[k_{z}] \ge 0$). We have also defined
\begin{align}
\Delta n(\omega,T_{1},T_{2})
&=n(\omega,T_{1})-n(\omega,T_{2})
=\frac{1}{2}\left(\coth\left[\frac{\hbar\omega}{2k_{\rm B}T_{1}}\right]-\coth\left[\frac{\hbar\omega}{2k_{\rm B}T_{2}}\right]\right)
\nonumber\\
&=\frac{1}{2}\tanh \left[\frac{\hbar\omega}{2k_{\rm B}}\left(\frac{1}{T_{1}}-\frac{1}{T_{2}}\right)\right]\left(1-\coth\left[\frac{\hbar\omega}{2k_{\rm B}T_{1}}\right]\coth\left[\frac{\hbar\omega}{2k_{\rm B}T_{2}}\right]\right)~,
\end{align}
where $n(\omega,T)=1/[e^{\frac{\hbar\omega}{k_{\rm B}T}}-1]$ is the Bose-Einstein occupation number.

As pointed out in the main text, the previous result allows for the consideration of multilayered structures. In this case, numbering the layers in the stack from the top ($n=1$ corresponds to the medium above the first interface) to the bottom, the reflection coefficients, as seen from an electromagnetic wave impinging from the top of the layer onto the top most interface, can be obtained using following recurrence formula ~\cite{Zhang05,Yariv83}
\begin{equation}
r^{\sigma}_{n}=\frac{\tilde{r}^{\sigma}_{n}+r^{\sigma}_{n+1}e^{-2 t_{n+1}\kappa_{n+1}}}{1+\tilde{r}^{\sigma}_{n}r^{\sigma}_{n+1}e^{-2 t_{n+1}\kappa_{n+1}}}~,
\end{equation}
where $\tilde{r}^{\sigma}_{n}$ is the interface reflection coefficient between the layer $n$ and $n+1$, $t_{n}$ is the thickness of the $n$th-layer and $\kappa_{n}=\sqrt{k^{2}-\epsilon_{n}(\omega)\omega^{2}/c^{2}}$, with $\epsilon_{n}(\omega)$ the corresponding permittivity. In case of a finite multilayer structure having $N$ layers, we set $r^{\sigma}_{N}=\tilde{r}^{\sigma}_{N}$. For local and isotropic materials the expression for $\tilde{r}^{\sigma}_{n}$ can be given in terms of Fresnel coefficients~\cite{Bimonte11b,Klimchitskaya:2019nzu}
\begin{align}
\tilde{r}^{TE}_{n}&=\frac{\sqrt{k^{2}-\epsilon_{n}(\omega)\frac{\omega^{2}}{c^{2}}}-\sqrt{k^{2}-\epsilon_{n+1}(\omega)\frac{\omega^{2}}{c^{2}}}}{\sqrt{k^{2}-\epsilon_{n}(\omega)\frac{\omega^{2}}{c^{2}}}+\sqrt{k^{2}-\epsilon_{n+1}(\omega)\frac{\omega^{2}}{c^{2}}}}~,
\nonumber\\
\tilde{r}^{TM}_{n}&=\frac{\epsilon_{n+1}(\omega)\sqrt{k^{2}-\epsilon_{n}(\omega)\frac{\omega^{2}}{c^{2}}}-\epsilon_{n}(\omega)\sqrt{k^{2}-\epsilon_{n+1}(\omega)\frac{\omega^{2}}{c^{2}}}}{\epsilon_{n+1}(\omega)\sqrt{k^{2}-\epsilon_{n}(\omega)\frac{\omega^{2}}{c^{2}}}+\epsilon_{n}(\omega)\sqrt{k^{2}-\epsilon_{n+1}(\omega)\frac{\omega^{2}}{c^{2}}}}~.
\end{align}

Commonly, multilayer structures are made out of metallic and insulating layers. One of the simplest mathematical expressions for the dielectric function of metals is given by the Drude model which was already presented in \eqnref{eq:drude}. For semiconductors or insulators a correspondingly simple description is given by the Lorentz model
\begin{equation}
\label{eq:lorentz}
\epsilon(\omega)=\epsilon_{\infty}+\frac{(\epsilon_{0}-\epsilon_{\infty})\Omega_{0}^{2}}{\Omega_{0}^{2}-\omega^{2}-{\rm i} \Gamma \omega}~.
\end{equation}
For example, this expression has been used in~\figref{fig:OoEPressure} to describe the optical properties for both silicon and silica. For simplicity, both materials were described using the parameters provided in Ref.~\cite{Pirozhenko08} to which we add a small dissipation rate to account for the material dissipation near resonance [see~\figref{fig:OoEPressure}].

\figref{fig:OoEPressure} also presents a calculation involving the pressure gradient $P'(a,T_{1},T_{2})$. 
Although the expression for $P'(a,T_{1},T_{2})$ can be obtained analytically from the expression for $P(a,T_{1},T_{2})$, the numerical evaluation of the corresponding result can
be quite unstable. For this reason, the pressure gradient was obtained by applying a symmetric eighth-order numerical differentiation algorithm which gives an estimate of the derivative of the function $f(x)$ at the point $x_{0}$ as
\begin{align}
f'(x_{0})\approx \frac{1}{\delta}&\left[\frac{1}{280}f(x_{0}-4\delta)-\frac{4}{105}f(x_{0}-3\delta)+\frac{1}{5}f(x_{0}-2\delta)-\frac{4}{5}f(x_{0}-\delta)\right.
\nonumber\\
&\left.+\frac{4}{5}f(x_{0}+\delta)-\frac{1}{5}f(x_{0}+2\delta)+\frac{4}{105}f(x_{0}+3\delta)-\frac{1}{280}f(x_{0}+4\delta)\right]~.
\end{align}
for sufficiently small $\delta$.
The result was checked against the corresponding expression for the derivative with respect to the distance of the Lifshitz formula in \eqnref{Elifshitz}, which can be obtained 
by using the following identity
\begin{equation}
\partial_{a}\frac{e^{- 2\kappa a}}{1-r^{\sigma}_{1}r^{\sigma}_{2}e^{-2\kappa a}}=-\frac{2 \kappa e^{- 2\kappa a}}{(1-r^{\sigma}_{1}r^{\sigma}_{2}e^{-2\kappa a})^{2}}~.
\end{equation}
The comparison successfully validated the numerical differentiation scheme with $\delta=1/8$ $\mu$m to the level of one part in a million.


\end{document}